\documentclass[aps,prl,reprint, superscriptaddress]{revtex4-2}
\usepackage{graphicx} 
\usepackage{dcolumn}% Align table columns on decimal point
\usepackage{amssymb}
\usepackage{amsmath}
\usepackage{fancyhdr}
\usepackage{bm}% bold math
\usepackage{times}
\usepackage{mhchem}
\usepackage[colorlinks,linkcolor=blue,anchorcolor=blue,citecolor=blue,urlcolor=blue]{hyperref}
\usepackage{epstopdf}
\usepackage[normalem]{ulem}

\UseRawInputEncoding
\begin{document}
    \title{Superconducting qubit based on altermagnets}
    \author{Xue-Feng Pan}
    \affiliation{Ministry of Education Key Laboratory for Nonequilibrium Synthesis and Modulation of Condensed Matter, Shaanxi Province Key Laboratory of Quantum Information and Quantum Optoelectronic Devices, School of Physics, Xi'an Jiaotong University, Xi'an 710049, China}
    \author{Xin-Lei Hei}
    \affiliation{Ministry of Education Key Laboratory for Nonequilibrium Synthesis and Modulation of Condensed Matter, Shaanxi Province Key Laboratory of Quantum Information and Quantum Optoelectronic Devices, School of Physics, Xi'an Jiaotong University, Xi'an 710049, China}
    \author{Franco Nori}
   \affiliation{Quantum Information Physics Theory Research Team, Center for Quantum Computing, RIKEN, Wakoshi, Saitama 351-0198, Japan}
    \author{Peng-Bo Li}
    \email{lipengbo@mail.xjtu.edu.cn}
    \affiliation{Ministry of Education Key Laboratory for Nonequilibrium Synthesis and Modulation of Condensed Matter, Shaanxi Province Key Laboratory of Quantum Information and Quantum Optoelectronic Devices, School of Physics, Xi'an Jiaotong University, Xi'an 710049, China}
    
    \date{\today}% It is always \today, today,

\begin{abstract}
	Altermagnets, characterized by vanishing net magnetization and momentum-dependent spin splitting, provide a promising platform for next-generation Josephson devices.
    Here, we exploit the Josephson effect in superconductor-altermagnet-superconductor junctions and show how to engineer prescribed current-phase relations by device design.
	Based on these programmable Josephson potentials utilizing altermagnetism, we propose a new class of superconducting qubits that combine large anharmonicity with enhanced robustness against decoherence via coherent two-Cooper-pair tunneling. We show that in the $2\phi$-junction regime, this kind of qubit is intrinsically protected against both charge and flux noise due to parity protection. 
	Magnetic flux can be used to precisely control the qubit and, under appropriate bias, this architecture further suppresses charge and flux noise. Our results establish altermagnets as a versatile platform for Josephson-potential engineering and open a new route toward high-performance superconducting qubits combining high coherence, large anharmonicity, and broad tunability.
\end{abstract}
\maketitle

\textit{Introduction.}---Superconducting qubits are a leading platform for quantum computing~\cite{2008ClarkeP10311042,2011YouP589597,2017GuP1102,2019KockumP703741,2020KjaergaardP369395,2021BlaisP2500525005,2021KwonP4110241102,2006YamashitaP132501132501,2006YouP1451014510,2006GrajcarP172505172505,2007KatoP172502172502,2007YouP104516104516,2008YouP4700147001,2015HoiP10451049}, with Josephson junctions providing the essential nonlinear element. For these qubits, the central challenge is to suppress decoherence induced by environmental noise. Existing approaches to reducing noise fall into three broad categories. 
The first is to shunt the Josephson junction with a large capacitor or inductor, which enhances the coherence of flux qubits~\cite{1999MooijP10361039,1999OrlandoP1539815413,2007YouP140515140515,2010SteffenP100502100502,2016YanP1296412964}, charge qubits~\cite{2007KochP4231942319,2008SchreierP180502180502,2011PaikP240501240501,2013BarendsP8050280502,2015LangeP127002127002,2015LarsenP127001127001,2018CasparisP915919,2021PlaceP17791779,2022WangP33,2025FeldsteinBofillP4409944099}, and
fluxonium~\cite{2018LinP150503150503,2019NguyenP4104141041,2022BaoP1050210502,2023SomoroffP267001267001,2009ManucharyanP113116}.
These strategies, however, generally entail trade-offs, such as reduced anharmonicity or increased circuit complexity. The second strategy is to engineer protected effective potentials using multijunction circuits based on conventional superconductors like the $0$-$\pi$ qubit~\cite{YouP,2013BrooksP5230652306,2018GroszkowskiP4305343053,2019PaoloP4300243002,2021GyenisP1033910339,2026KolesnikowP1030610306} and $\cos(2\phi)$-dominated circuits~\cite{1996ZapataP22922295,2004SavelevP179179,2004SavelevP6610966109,2004Savel’evP403408,2020SmithP88,2025HaysP4032140321,RoverchP,ZhurbinaP}. Such schemes, nevertheless, require multiple Josephson elements, leading to increased circuit complexity and susceptibility to fabrication imperfections.
The third route exploits the intrinsically nonsinusoidal current-phase relations (CPRs) of unconventional superconducting junctions to generate higher Josephson harmonics,  such as the $d$-wave-based flowermon~\cite{2024BroscoP1700317003} and $d$-mon~\cite{2024PatelP1700217002}.
However, low-energy quasiparticles associated with the nodal superconducting gap can introduce additional dissipation.
These limitations thus call for a simpler platform combining higher-harmonic Josephson potentials with intrinsic noise protection, free from complex multijunction circuits and nodal unconventional superconductors.

Recently, altermagnets have emerged as a novel class of magnetic phases~\cite{2019NakaP43054305,2020SmejkalP88098809,2022BaiP197202197202,2022FengP735743,2022SmejkalP482496,2022ifmmodeSelseSfimejkalP4050140501,2022ifmmodeSelseSfimejkalP3104231042,2025SongP473485,2025TamangP,2026ChenP4670546705,2026LiP4670446704,2026LiuP869873} with profound implications for spintronics~\cite{2021ShaoP70617061,2022SmejkalP1102811028,2022BoseP267274,2022KarubeP137201137201,2024ZhangP23133322313332,2024BaiP24093272409327,2025FuP111111,2025HuangP266701266701,2025GuoP25057792505779,2026SudoP1650316503,2026JungwirthP10121021,2026MonkmanP1105711057,ZhaoP}, magnonics~\cite{2023ifmmodeSelseSfimejkalP256703256703,2023JinP9092307,2025CostaP125125,2025GarciaGaitanP2040720407,2025BeidaP9797,2025SourounisP134448134448,2025EtoP9444294442,2025BiniskosP93119311,2025JinP126702126702,2026JinP8670386703,2026WeissenhoferP2525,2026YuanP106901106901,2026WiedmannP2626,2026RodriguezSuarezP134411134411,2026YangP2670126701,2026LiP224403224403}, superconductivity~\cite{2023PapajP6050860508,2023SunP5451154511,2023OuassouP7600376003,2023BeenakkerP7542575425,2024DasP245424245424,2024ChengP1451814518,2024ZhangP18011801,2024BanerjeeP2450324503,2024LuP226002226002,2025SumitaP144510144510,2025MazinP1818,2026JasiewiczP5151,2026LiuP}, and topological physics~\cite{2023LiP205410205410,2024GhorashiP106601106601,2024LiP201109201109,2025LiuP241405241405,2026HodgeP4545,2026FuP9660496604,2026YangP4545,2026HuoP2442024420,2026GonzalezGarciaP4400444004}.
Altermagnets possess a collinear compensated magnetic order and exhibit sizable spin splitting in momentum space. They
feature spin-dependent band structures analogous to those of ferromagnets but maintain a zero net magnetization~\cite{2022ifmmodeSelseSfimejkalP4050140501,2022ifmmodeSelseSfimejkalP3104231042,2025SongP473485,2025TamangP}.
Consequently, they effectively mitigate the detrimental stray fields associated with conventional magnets that typically suppress superconductivity~\cite{2026YangP4545,2004GolubovicP546546}. Meanwhile, they introduce additional degrees of freedom to manipulate transport properties in Josephson junctions.
Furthermore, altermagnet-superconductor heterostructures exhibit a rich variety of unconventional superconducting phenomena~\cite{2025ChakrabortyP2600126001,2025M.FroldiP170273170273,2025FukayaP6450264502,2025AlipourzadehP214515214515,2025FukayaP313003313003,2025MaedaP144508144508,VosoughiniaP,2026LuP180501180501,2026FukayaP226001226001}.
This development opens up promising avenues for engineering anomalous Josephson potentials.
In particular, altermagnetic Josephson junctions exhibit orientation-dependent Andreev reflection~\cite{2023PapajP6050860508,2023SunP5451154511}, $0$-$\pi$ transitions~\cite{2023OuassouP7600376003,2024LuP226002226002}, and multi-harmonic CPRs~\cite{2024LuP226002226002,2024ChengP2451724517,2025SunP165406165406,2025ZhaoP184515184515}.
Despite these advances, current research on altermagnetic Josephson junctions has predominantly focused on fundamental transport phenomena, leaving their potential for superconducting qubit and quantum device applications largely unexplored.

In this work, we propose a superconducting-qubit architecture based on altermagnetic Josephson junctions (AMJJs).
We first analyze and show how to engineer nonsinusoidal Josephson potentials of an AMJJ by designing the microscopic parameters of the junction. 
Within a model retaining only single- and double-Cooper-pair tunneling processes, the AMJJ can be tuned into either the $\phi$- or the $2\phi$-junction regime~\cite{2007GoldobinP224523224523,2012SickingerP107002107002,2025MitrovicP6700167001}.
We then harness AMJJs to construct superconducting qubits and investigate their quantum properties across different junction regimes.
We show that this kind of qubit has several advantages
compared to previous superconducting qubits.
It has \textit{strong anharmonicity} in both the $\phi$- and $2\phi$-junction regimes.
It also provides \textit{intrinsic protection against both charge and flux noise}. In the $2\phi$-junction regime, charge and flux noise are completely suppressed due to the parity of the qubit wavefunctions~\cite{SM,KrutiP}, in direct contrast to previous $d$-wave superconducting qubits~\cite{2024BroscoP1700317003,2024PatelP1700217002} and other proposals~\cite{YouP,2007KochP4231942319,2007YouP140515140515,2009ManucharyanP113116,2013BrooksP5230652306}.
These noise-protection properties persist over a wide range of tunable parameters, highlighting the flexibility of AMJJs for superconducting-qubit design.
Furthermore, $s$-wave superconductors of AMJJs with an isotropic gap suppress quasiparticle noise while remaining compatible with established superconducting-circuit fabrication techniques~\cite{2011CatelaniP7700277002,SM}.

To facilitate qubit control and readout, we further investigate a flux-tunable implementation of the proposed architecture.
Tuning the external magnetic flux can control the contribution of double-Cooper-pair tunneling and suppress sensitivity to charge and flux noise.
%The contribution of double-Cooper-pair tunneling can be controlled using an external magnetic flux.
%Selecting appropriate operating points by tuning the external flux allows complete suppression of charge-noise sensitivity.
%Although double-Cooper-pair tunneling generally enhances flux-noise sensitivity, this effect can be suppressed by choosing an appropriate flux-bias point.
This work establishes altermagnets as a versatile and programmable platform for Josephson-potential engineering, and provides a promising route toward high-performance superconducting qubits with strong anharmonicity, intrinsic noise protection, and flexible control.

\begin{figure}
	\centering
	\includegraphics[width=0.48\textwidth]{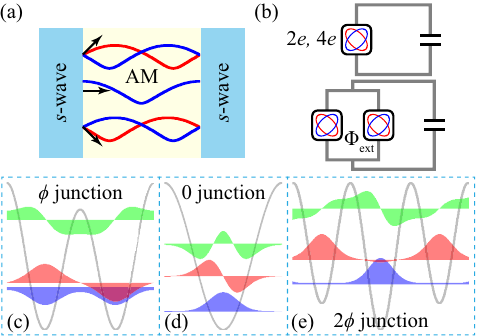}
	\caption{(a) Schematic of an AMJJ, consisting of an altermagnetic layer of thickness $L$ sandwiched between two $s$-wave superconductors. The red (blue) curves denote spin-up (spin-down). (b) Schematic of a superconducting qubit circuit based on the AMJJ. Panels (c-e) schematically illustrate the Josephson energy corresponding to the $\phi$-, $0$-, and $2\phi$-junction realized with the AMJJ, respectively. The blue, red, and yellow shaded regions represent the wave-function distributions of the three lowest-energy eigenstates, respectively, while their vertical positions indicate the corresponding eigenenergies. The ground and first excited states chosen as the qubit states $\vert 0\rangle$ and $\vert 1\rangle$.}
	\label{FIG1}
\end{figure}

\textit{Tailoring of the Josephson energy.}---We employ an AMJJ as a nonlinear element for realizing superconducting qubits, as schematically illustrated in Figs.~\ref{FIG1}(a,~b).
Specifically, the AMJJ enables the on-demand synthesis of non-sinusoidal Josephson potentials containing higher-harmonic terms associated with multiple-Cooper-pair tunneling processes.
Figures~\ref{FIG1}(c-e) show the Josephson potentials of a $\phi$-junction, a $0$-junction, and a $2\phi$-junction, respectively.
As illustrated by these examples, higher-order Cooper-pair tunneling processes can enhance the anharmonicity of the spectrum.

To elucidate how the microscopic parameters shape the Josephson potential, we now formulate the microscopic Hamiltonian of the AMJJ.
The AMJJ features a trilayer structure consisting of two semi-infinite $s$-wave superconducting leads separated by an altermagnetic interlayer.
For simplicity, we focus on a $d_{xy}$-wave altermagnet ($d_{xy}$-AM; the general case is provided in the Supplementary Material~\cite{SM}). 
Using the Nambu spinor $\hat{c}_\mathbf{k}=[\hat{c}_{\mathbf{k},\uparrow},\hat{c}_{\mathbf{k},\downarrow},\hat{c}_{-\mathbf{k},\uparrow}^\dagger,\hat{c}_{-\mathbf{k},\downarrow}^\dagger]$, the Hamiltonian of the altermagnetic region is expressed as $\hat{H}_\mathrm{AM}=1/2\sum_\mathbf{k}\hat{c}_\mathbf{k}^\dagger\hat{\mathcal{M}}_\mathbf{k}\hat{c}_\mathbf{k}$, where $\hat{\mathcal{M}}_k=\alpha_1 k_x k_y\hat{s}_z\hat{\tau}_z$ with $\mathbf{k}=(k_x,k_y)$ and $\alpha_1=2J\sin(2\alpha)/k_F^2$~\cite{2022SmejkalP1102811028,SM}.
Here, $J$ denotes the exchange energy, $\alpha$ represents the misorientation angle between the altermagnetic crystalline axis and the interface normal, $k_F$ is the Fermi wave vector, and $\hat{s}_z$ ($\hat{\tau}_z$) is the Pauli matrix acting on the spin space (Nambu space).
We focus here on $\alpha=\pi/4$, where the CPR is insensitive to small deviations in the crystal-axis orientation to first order (see the Supplemental Material~\cite{SM} for a detailed proof).

We focus on the short-junction limit ($L\ll\xi_0$, where $\xi_0$ denotes the superconducting coherence length), in which the Josephson current is dominated by contributions from Andreev bound states~\cite{2023BeenakkerP7542575425,1991BeenakkerP38363839,1999FurusakiP809818}.
Altermagnetism lifts the spin degeneracy of these states, giving rise to the following spin-resolved branches [three positive branches are sketched in Fig.~\ref{FIG1}(a); red/blue denote spin up/down]:
\begin{equation}
 E_\uparrow^\pm(k_y)=\pm\Delta(T)\sqrt{1-T(k_y)\sin^2(\phi/2-\alpha_\mathrm{AM})}   
\end{equation}
and
\begin{equation}
E_\downarrow^\pm(k_y)=\pm\Delta(T)\sqrt{1-T(k_y)\sin^2(\phi/2+\alpha_\mathrm{AM})}.
\end{equation}
Here, $\alpha_\mathrm{AM}=k_yJL$ denotes the altermagnet-induced phase shift.
The interfacial barrier parameter $Z=mU_I/(\hbar^2 k_F)$ and the chemical-potential mismatch $\delta\!\mu$ control the momentum-dependent transmission probability $T(k_y)$, where $m$ is the electron mass and $U_I$ denotes the interfacial barrier height.
Each momentum channel can therefore be viewed as an effective Josephson junction with its own phase shift and transmission probability, and the macroscopic Josephson potential is obtained by summing over all channels.
In this picture, $J$ and $L$ tune the channel-dependent phase shifts, whereas $Z$ and $\delta\!\mu$ regulate the relative channel contributions.
Detailed numerical results illustrating these distinct roles are provided in the Supplemental Material~\cite{SM}.

\begin{figure}
	\centering
	\includegraphics[width=0.48\textwidth]{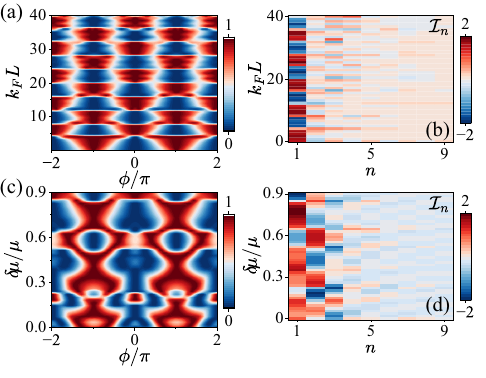}
	\caption{(a,~c) Modulation of the AMJJ Josephson energy by the altermagnet thickness $L$ and a chemical-potential mismatch $\delta\!\mu$, respectively. (b,~d) Corresponding harmonic components of the Josephson current for the cases shown in panels (a) and (c), respectively. The fixed parameters are $J=0.4\mu$, $\delta\mu=0$, and $Z=0$ in panels (a,~b), and $J=0.4\mu$, $k_FL=20$, and $Z=0$ in panels (c,~d).}
	\label{FIG2}
\end{figure}

Having established a microscopic picture of quantum interference through the Andreev-bound-state analysis, we next numerically evaluate the total Josephson current $I(\phi)$ using the Furusaki-Tsukada formula~\cite{2021AsanoP,2000KashiwayaP16411641,1991FurusakiP299302}.
The current is expressed as $I(\phi) = I_c\mathcal{I}(\phi)$, where $I_c$ is the critical current and $\mathcal{I}(\phi)$ is the normalized CPR. The corresponding macroscopic Josephson potential is thus given by $\mathcal{V}=E_J\mathcal{U}(\phi)$, with the Josephson energy $E_J=\Phi_0I_c/(2\pi)$, the dimensionless Josephson potential $\mathcal{U}(\phi)=\int_{0}^\phi\mathcal{I}(\phi^\prime)d\phi^\prime$, and the flux quantum $\Phi_0$.
Figure~\ref{FIG2}(a) shows that varying the altermagnetic interlayer thickness $L$ primarily induces $0\text{-}\pi$ transitions, whereas $\phi$ and $2\phi$ junctions are realized only within exceedingly narrow ranges of $L$.
A Fourier analysis of the CPR [Fig.~\ref{FIG2}(b)] further shows that the first harmonic dominates over most of the $L$ range, while higher harmonics become appreciable only near the $0\text{-}\pi$ transition boundaries.
The results presented in Figs.~\ref{FIG2}(c,~d) demonstrate that a finite $\delta\!\mu$ not only broadens the parameter regions supporting $\phi$ and $2\phi$ junctions but also substantially enhances the amplitudes of the higher harmonics.
The effects of the exchange interaction strength $J$ and the interfacial barrier parameter $Z$ are analogous to those of the altermagnet thickness $L$ and the chemical-potential mismatch $\delta\!\mu$, respectively, as detailed in the Supplemental Material~\cite{SM}.

\textit{AM-superconducting qubit.}---Accounting for both single- and double-Cooper-pair tunneling, the CPR is expressed as
\begin{equation}
\mathcal{I}(\phi)=\mathcal{I}_1\sin(\phi)+\mathcal{I}_2\sin(2\phi),    
\end{equation}
with the corresponding microscopic parameters determined via an inverse-design procedure~\cite{SM}.
As illustrated in Fig.~\ref{FIG1}(b), we consider shunting the AMJJ with a large capacitor to form a superconducting qubit, whose Hamiltonian reads
\begin{equation}
	\hat{H}_\mathrm{Tr}=4E_C(\hat{N}-N_g)^2-E_J\mathcal{I}_1\cos(\hat{\phi})-E_J\frac{\mathcal{I}_2}{2}\cos(2\hat{\phi}).
	\label{H_Tr}
\end{equation}
Here, $E_C$ denotes the charging energy, $N_g$ is the offset charge (we assume $N_g=0$ for simplicity), and $\hat{N}=-i\partial_\phi$ represents the Cooper-pair number operator.

\begin{figure}
	\centering
	\includegraphics[width=0.5\textwidth]{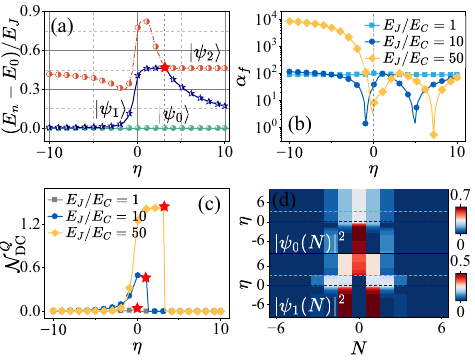}
	\caption{(a) Eigenenergy spectrum of the superconducting qubit as a function of $\eta$ for $E_J/E_C=50$. The red star shows the critical value $\eta_c$, at which the first and second excited states become degenerate. (b) Anharmonicity $\alpha_f$ and (c) transverse charge-noise coupling strength $\mathcal{N}_\mathrm{DC}^Q$ versus $\eta$ for different $E_J/E_C$. The stars in (c) indicate the corresponding critical values $\eta_c$ for different $E_J/E_C$. (d) Wave functions of the ground and first excited states in the Cooper-pair-number ($\hat N$) basis for $E_J/E_C=50$. The blue and green dashed curves correspond to $\eta=0$ and $\eta=\eta_c$, respectively.}
	\label{FIG3}
\end{figure}

To characterize the influence of higher-order harmonics on the Josephson potential, we introduce the dimensionless ratio $\eta = \mathcal{I}_2/\mathcal{I}_1$.
As illustrated in Fig.~\ref{FIG1}(c-e), the regimes $\eta<0$, $\eta=0$, and $\eta>0$ correspond to the $\phi$-junction, $0$-junction, and $2\phi$-junction regimes, respectively.
A larger value of $\vert\eta\vert$ indicates a stronger relative contribution from two-Cooper-pair tunneling.
Figure~\ref{FIG3}(a) shows the dependence of the superconducting-qubit energy spectrum on $\eta$.
In the $\eta<0$ ($\phi$-junction) regime, the ground and first excited states become nearly degenerate as $\eta$ decreases.
By contrast, in the $\eta>0$  ($2\phi$-junction)  regime, the first and second excited states become degenerate at a critical value of $\eta$, denoted by $\eta_c$ [the red stars in Fig.~\ref{FIG3}(a,~c)].
Upon crossing $\eta_c$, the parity of the first excited state changes, as illustrated in Figs.~\ref{FIG1}(c-e) and further discussed in the Supplemental Material~\cite{SM}.
%Beyond this critical point, the qubit’s sensitivity to charge noise is completely suppressed, as discussed in detail below.
Figure~\ref{FIG3}(b) shows the dependence of the relative qubit anharmonicity $\alpha_f=\vert f_{12}-f_{01}\vert / f_{01}\times 100 \%$ on $\eta$, where $f_{01}$ and $f_{12}$ are the corresponding transition frequencies.
For $0$-junction superconducting qubits, insensitivity to low-frequency charge noise is typically achieved by operating deep in the regime, $E_J/E_C\gg1$~\cite{2007KochP4231942319,2021RasmussenP4020440204}.
However, this suppression of low-frequency charge-noise sensitivity comes at the cost of reduced anharmonicity, as indicated by the intersections of the gray dashed line ($\eta=0$) with the individual curves in Fig.~\ref{FIG3}(b).
In contrast, here, the qubit anharmonicity can be enhanced by tuning $\eta$, despite operating when $E_J/E_C\gg1$.

We now proceed to investigate the impact of environmental noise on qubit coherence.
% Within our model, we focus primarily on the effects of charge noise. The effect of flux noise is detailed in the Supplemental Material~\cite{SM}.
We focus primarily on charge, flux, and quasiparticle noise.
In our model, the latter two exhibit the same protection properties as charge noise.
We therefore present only the charge-noise results in the main text.
Analyses of the other noise sources are given in the Supplemental Material~\cite{SM}.
We first consider energy relaxation arising from high-frequency charge noise. According to Fermi's golden rule, the relaxation rate can be expressed as~\cite{SM}
\begin{equation}
	\Gamma_\mathrm{DC}^Q = \frac{1}{\hbar^2}64E_C^2\mathcal{N}_\mathrm{DC}^Q S_Q(\omega_q),
	\label{Gamma_DC_Q}
\end{equation}
where $\mathcal{N}_\mathrm{DC}^Q=\vert \langle \psi_0\vert\hat{N}\vert \psi_1
\rangle\vert^2$ denotes the transverse charge-noise coupling strength, and $S_Q(\omega_q)$ is the charge-noise spectral density evaluated at the qubit transition frequency $\omega_q=2\pi f_{01}$.
Thus, $\mathcal{N}_\mathrm{DC}^Q$ directly determines the qubit's sensitivity to high-frequency charge noise.
As shown in Fig.~\ref{FIG3}(c), for $\eta<0$, the transverse charge-noise coupling decreases monotonically with decreasing $\eta$, thereby suppressing energy relaxation.
For $\eta>0$, however, the transverse charge-noise coupling initially increases with $\eta$, indicating enhanced sensitivity to charge noise. Upon reaching the critical point $\eta_c$, the transverse charge-noise coupling drops abruptly to a value substantially smaller than those observed in the $\eta<0$ regime, signifying a sharp suppression of charge-noise sensitivity.

Next, we elucidate this phenomenon from the parity of the eigenstates.
We know that the operator $\hat{N}=-i\partial_\phi$ is an odd-parity operator.
For $\eta<\eta_c$, the ground and first excited states possess opposite parity [see Fig.~\ref{FIG1}(c) and the Supplemental Material~\cite{SM}].
Consequently, $\mathcal{N}_\mathrm{DC}^Q$ is governed by the charge-operator matrix element between these two states, whose magnitude depends on the spatial overlap of their wave functions.
For $\eta<0$, the state overlap is small [Figs.~\ref{FIG3}(d)], leading to a small $\mathcal{N}_\mathrm{DC}^Q$.
When $0 < \eta < \eta_c$, the state overlap increases, and thus $\mathcal{N}_\mathrm{DC}^Q$ increases.
In the regime $\eta > \eta_c$, the ground and first excited states have the same parity [see Fig.~\ref{FIG1}(e)].
Because the charge-number operator $\hat{N}$ has odd parity, the corresponding parity selection rule enforces $\vert \langle \psi_0\vert\hat{N}\vert \psi_1\rangle\vert^2=0$ throughout this regime, thereby completely suppressing the associated charge-noise-induced decay.
Notably, the critical value $\eta_c$ shifts progressively toward larger values as $E_J/E_C$ increases [see Fig.~\ref{FIG3}(c) and the Supplemental Material~\cite{SM}].
Thus, two-Cooper-pair tunnelling enables strong suppression of charge-noise-induced transitions while retaining substantial anharmonicity at moderate $E_J/E_C$.

\begin{figure}
	\centering
	\includegraphics[width=0.5\textwidth]{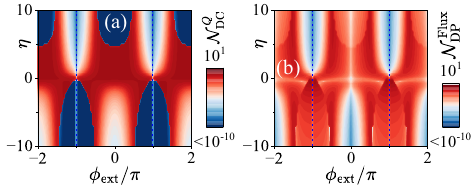}
	\caption{(a) transverse charge-noise coupling strength $\mathcal{N}_\mathrm{DC}^Q$, and (b) longitudinal flux-noise coupling strength $\mathcal{N}_\mathrm{DP}^\mathrm{Flux}$ versus the external flux $\phi_\mathrm{ext}$ and the ratio $\eta$. The fixed parameters are $E_J/E_C=50$ and $N_g=0$.}
	\label{FIG4}
\end{figure}

The pure-dephasing rate induced by low-frequency charge noise is given by~\cite{SM,2019KrantzP2131821318,2021RasmussenP4020440204,2001MakhlinP357400}
\begin{equation}
\Gamma_\mathrm{DP}^Q=\frac{1}{\hbar^2}64E_C^2\mathcal{N}_\mathrm{DP}^QS_Q(0). 
\end{equation}
Equivalently, we define the longitudinal charge-noise coupling strength as $\mathcal{N}_\mathrm{DP}^Q=\vert \langle \psi_0\vert\hat{N}\vert \psi_0
\rangle-\langle \psi_1\vert\hat{N}\vert \psi_1
\rangle\vert^2$, where $S_Q(0)$ denotes the charge-noise spectral density at zero frequency.
Similarly, according to the parity properties of the states and the operator, $\langle\psi_i\vert\hat{N}\vert\psi_i\rangle=0$ holds constantly, meaning that first-order pure dephasing induced by low-frequency charge noise can be completely suppressed through parity protection.

\textit{Flux-tuned AM-superconducting qubit.}---By constructing a symmetric SQUID comprising two identical, symmetric AMJJs, we can achieve flux-tunable control and readout of the qubit.
The corresponding Hamiltonian reads 
\begin{equation}
 \hat{H}_\mathrm{FTr}=4E_C(\hat{N}-N_g)^2+\mathcal{U}_\mathrm{JJ}^\mathrm{eff},   
\end{equation}
where the effective Josephson potential is given by~\cite{SM}
\begin{equation}
	\mathcal{U}_\mathrm{JJ}^\mathrm{eff}=E_J\left[-2\mathcal{I}_1\cos\left(\frac{\phi_\mathrm{ext}}{2}\right)\cos(\phi)-\mathcal{I}_2\cos(\phi_\mathrm{ext}) \cos(2\phi)\right].
\end{equation}
As can be seen from the effective Josephson potential $\mathcal{U}_\mathrm{JJ}^\mathrm{eff}$, varying the external flux $\phi_\mathrm{ext}$ enables the potential to be tuned between cosine and noncosine forms.
% By contrast, for a conventional Josephson junction containing only the first harmonic $\mathcal{I}_1$, the applied flux cannot reshape the Josephson potential.
The flux tunability of the potential allows both the qubit frequency and anharmonicity to be controlled~\cite{SM}.

We next investigate the qubit coherence under flux control.
We consider only energy relaxation induced by high-frequency charge noise, with the corresponding relaxation rate given by Eq.~(\ref{Gamma_DC_Q}).
Figure~\ref{FIG4}(a) shows that as $\vert\eta\vert$ increases, charge noise is suppressed, as discussed previously.
In addition, the critical value $\eta_c$ for complete suppression of charge noise can be tuned by $\phi_\mathrm{ext}$.
Notably, when the external flux biases the qubit at the half-flux points $\Phi_\mathrm{ext}/\Phi_0=\pm1/2$, $\eta_c$ is tuned close to $0$.
In other words, charge noise is suppressed throughout the region $\vert \eta\vert\neq0$.
This is because at this operating point, SQUID interference completely cancels the first-harmonic contribution, leaving the effective Josephson potential dominated by the second harmonic, $\mathcal{U}_\mathrm{JJ}^\mathrm{eff} \propto \cos(2\phi)$.

Due to the SQUID structure in the circuit, qubits are susceptible to flux noise, which typically induces qubit transition-frequency fluctuations, thereby causing pure dephasing.
The flux-noise-induced pure-dephasing rate can be expressed as~\cite{SM,2019KrantzP2131821318,2021RasmussenP4020440204,2001MakhlinP357400}
\begin{equation}
	\Gamma_\mathrm{DP}^\mathrm{Flux}=\frac{1}{\hbar^2}E_J^2\mathcal{N}_\mathrm{DP}^\mathrm{Flux}S_\mathrm{Flux}(0),
\end{equation}
where $\mathcal{N}_\mathrm{DP}^\mathrm{Flux}$ denotes the longitudinal coupling strength to flux noise~\cite{ScrNDPFlux}, and $S_\mathrm{Flux}(0)$ is the zero-frequency flux-noise spectral density.
As shown in Fig.~\ref{FIG4}(b), the presence of higher-order harmonics increases the sensitivity of the flux-tunable superconducting qubit to flux noise to some extent.
At $\phi_\mathrm{ext}=(2n+1)\pi$ (with $n$ an integer), the first harmonic in the Hamiltonian vanishes, while the second harmonic remains.
Numerical results show that $\mathcal N_{\rm DP}^{\rm Flux}$ decreases near these points.
Notably, $\mathcal{N}_\mathrm{DP}^\mathrm{Flux}$ vanishes at $\phi_\mathrm{ext}=2n\pi$. These points are therefore first-order zero-coupling points for longitudinal flux noise~\cite{ScrNDPFlux}.

\textit{Conclusion.}---In this work, we first investigate how the microscopic structural parameters of AMJJs determine their Josephson potentials and, combined with an inverse-design approach, achieve the targeted design of device parameters from a prescribed CPR.
Subsequently, we consider AMJJs in which single- and double-Cooper-pair tunneling processes are dominant.
Based on this, we propose a superconducting qubit and show that as the contribution of the double-Cooper-pair tunneling process increases, charge-noise-induced decay can be suppressed.
In particular, for a $2\phi$-junction, charge and flux noise can be completely suppressed due to the parity properties of the states.
By incorporating a SQUID structure, we can realize flux control and readout of the qubit.
Although the introduction of higher-order processes leads to increased sensitivity to flux noise, this sensitivity can be mitigated by operating the qubit at specific bias points.
%In the protected regime, $\langle \psi_0\vert\hat{N}\vert\psi_1\rangle$ is suppressed in the $\phi$-junction regime and vanishes for $\eta>\eta_c$ in the $2\phi$-junction regime, limiting direct capacitive driving.
%For a single-junction qubit, higher excited states may assist indirect control and resonator-based dispersive readout~\cite{2024BroscoP1700317003}.
%In the SQUID implementation, external flux tunes the capacitive and flux couplings, providing channels for qubit control and dispersive readout\cite{2024PatelP1700217002}.
The SQUID implementation also offers a possible route to qubit control and dispersive readout through flux-tunable couplings.
Since AMJJs are implemented based on $s$-wave superconductors, they naturally suppress quasiparticle noise and are compatible with existing fabrication technologies~\cite{EsinP,SapkotaP}.
This work highlights the potential of AMJJs as a programmable platform for Josephson-potential engineering, providing a new route toward superconducting qubits that combine large anharmonicity, long coherence times, and strong tunability.

\textit{Note added.}---—During the completion of this work, we became aware of related independent studies~\cite{2026BratlandTjernshaugenP,2026GuoP}.

\begin{acknowledgments}
The authors are grateful to Max Hays, Jeonghun Sohn, and Daniele Lamberto for their valuable suggestions and helpful discussions.
XFP is supported by the National Natural Science Foundation of China under Grant Nos.~12604548 and 124B2091, the Fundamental Research Funds for the Central Universities under Grant No.~xzy012026058, and the China Postdoctoral Science Foundation under Grant No.~2026M793735.
PBL is supported by the National Natural Science Foundation of China under Grants No.~W2411002 and No.~12375018. 
XLH is supported by the National Natural Science Foundation of China (No.~12505029), the Fundamental Research Funds for the Central Universities of Ministry of Education of China (No.~xzy012025077), and the China Postdoctoral Science Foundation (No.~2025M773347).
FN is supported in part by the Japan Science and Technology Agency (JST) [via
the CREST Quantum Frontiers program Grant No.~JPMJCR24I2, the Quantum Leap Flagship Program (Q-LEAP), the Moonshot R\&D Grant No.~JPMJMS256E, and the ASPIRE program (Grant No.~JPMJAP2513)].
\end{acknowledgments}
    % \nocite{*}
% \bibliographystyle{apsrev4-2}
\bibliography{ArticleReference}
%\clearpage
\end{document}

\title{Supplemental Material for ``Superconducting qubit based on altermagnets''}

\author{Xue-Feng Pan}
\affiliation{Ministry of Education Key Laboratory for Nonequilibrium Synthesis and Modulation of Condensed Matter, Shaanxi Province Key Laboratory of Quantum Information and Quantum Optoelectronic Devices, School of Physics, Xi'an Jiaotong University, Xi'an 710049, China}
\author{Xin-Lei Hei}
\affiliation{Ministry of Education Key Laboratory for Nonequilibrium Synthesis and Modulation of Condensed Matter, Shaanxi Province Key Laboratory of Quantum Information and Quantum Optoelectronic Devices, School of Physics, Xi'an Jiaotong University, Xi'an 710049, China}
 \author{Franco Nori}
   \affiliation{Quantum Information Physics Theory Research Team, Center for Quantum Computing, RIKEN, Wakoshi, Saitama 351-0198, Japan}
\author{Peng-Bo Li}
\email{lipengbo@mail.xjtu.edu.cn}
\affiliation{Ministry of Education Key Laboratory for Nonequilibrium Synthesis and Modulation of Condensed Matter, Shaanxi Province Key Laboratory of Quantum Information and Quantum Optoelectronic Devices, School of Physics, Xi'an Jiaotong University, Xi'an 710049, China}


\begin{abstract}
In this Supplemental Material, we provide calculation details not covered in the main text. In Sec.~\ref{MMAMJJ}, we present a detailed microscopic model and an inverse-design scheme for the altermagnetic Josephson junction. Detailed derivations of the Hamiltonians and coherence analyses for the AM superconducting qubit and the flux-tunable AM superconducting qubit are provided in Sec.~\ref{AMSQ} and Sec.~\ref{FTAMSQ}, respectively.
\end{abstract}
\maketitle

% \newpage
\tableofcontents
\newpage

\section{\label{MMAMJJ}Microscopic Model of an Altermagnetic Josephson Junction}
In this section, we present a microscopic description of the Josephson junction.
As illustrated in Fig.~1(a) of the main text, the junction has an superconductor-altermagnet-superconductor (S-AM-S) sandwich geometry, in which both superconductors are characterized by $s$-wave pairing symmetry.
The altermagnet is located in the region $0<x<L$, while the left and right superconductors are located in the regions $x<0$ and $x>L$, respectively.
Here, $L$ denotes the thickness of the altermagnet.
In the Nambu basis defined by the spinor $\hat{c}_\mathbf{k}=[\hat{c}_{\mathbf{k},\uparrow},\hat{c}_{\mathbf{k},\downarrow},\hat{c}_{-\mathbf{k},\uparrow}^\dagger,\hat{c}_{-\mathbf{k},\downarrow}^\dagger]^T$, the Bogoliubov-de Gennes (BdG) Hamiltonian of the S-AM-S Josephson junction can be written as
\begin{equation}
	\hat{H}=\frac{1}{2}\sum_\mathbf{k}\hat{c}_\mathbf{k}^\dagger\hat{\mathcal{H}}_\mathbf{k}\hat{c}_\mathbf{k},
\end{equation}
where
\begin{equation}
	\hat{\mathcal{H}}_\mathbf{k}=\left(\frac{\hbar\mathbf{k}^2}{2m}-\widetilde{\mu}+U\right)\hat{\tau}_z+M_\mathbf{k}\hat{\tau}_z\otimes\hat{s}_z+i\left(\widetilde{\Delta}\hat{\tau}_+-\widetilde{\Delta}^*\hat{\tau}_-\right)\otimes\hat{s}_y.
	\label{H_S_AM_S}
\end{equation}
Here, $\hat{\tau}_i$ and $\hat{s}_i$ denote the Pauli matrices acting in Nambu and spin spaces, respectively.

The first term of the Eq.~\eqref{H_S_AM_S} consists of three contributions: $\hbar\mathbf{k}^2/2m$ denotes the electron kinetic energy; $\widetilde{\mu}=\mu+\delta\mu\Theta(x)\Theta(L-x)$ describes the spatially dependent chemical potential across the junction; and $U=U_I[\delta(x)+\delta(x-L)]$ represents the $\delta$-function barriers at the two interfaces ($x=0$ and $x=L$).
Here, $\delta(x)$ and $\Theta(x)$ denote the Dirac delta function and the Heaviside step function, respectively.
Here, $\hbar$, $m$, $\mu$, $\delta\mu$, and $U_I$ denote the reduced Planck constant, the electron mass, the chemical potential in the S regions, the chemical-potential mismatch between the AM and S regions, and the strength of the interfacial $\delta$-function barriers, respectively.
For convenience, we introduce the dimensionless interfacial barrier strength $Z=mU_I/(\hbar^2k_F)$, where $k_F$ denotes the Fermi wave vector.

The second term of the Eq.~\eqref{H_S_AM_S} describes the altermagnetic exchange interaction in the AM region and takes the form~\cite{2022SmejkalP1102811028}
\begin{equation}
	M_\mathbf{k}=\frac{\mathcal{J}}{k_F^2}\left[\left(k_x^2-k_y^2\right)\cos\left(2\alpha\right)+2k_xk_y\sin\left(2\alpha\right)\right],
\end{equation}
where $\mathcal{J}=J\Theta(x)\Theta(L-x)$ specifies the spatial profile of the exchange interaction, restricting it to the AM region ($0<x<L$).
Here, $J$ denotes the strength of the altermagnetic exchange interaction.
The parameter $\alpha$ denotes the angle between the lobe axis of the altermagnetic order and the interface normal.

The third term of the Eq.~\eqref{H_S_AM_S}, describing the $s$-wave superconducting pairing potential, is given by
\begin{equation}
	\widetilde{\Delta}=\Delta_T\left[e^{i\phi}\Theta(-x)+\Theta(x-L)\right].
\end{equation}
Here, we consider the temperature-dependent pairing potential $\Delta_T=\Delta_0\tanh(1.74\sqrt{T_c/T-1})$, where $\Delta_0$ is the amplitude of the pair potential at zero temperature, $T_c$ is the superconducting critical temperature, and $T$ is the system temperature.
Solving the BdG equation associated with the Eq.~\eqref{H_S_AM_S} yields the quasiparticle wave functions in each region of the junction, which are subsequently used to analyze its Josephson transport properties.

\subsection{\label{ES_BdG_Eq}Eigenmodes in Each Region}
To analyze the Josephson transport, we first determine the quasiparticle eigenstates in each region of the junction.
We begin with the left superconducting region $x<0$, where $\widetilde{\mu}=\mu$, $M_\mathbf{k}=0$, and $\widetilde{\Delta}=e^{i\phi}\Delta_T$.
Substituting these parameters into Eq.~\eqref{H_S_AM_S}, we obtain the BdG Hamiltonian in the left superconducting region as
\begin{equation}
	\hat{\mathcal{H}}_\mathbf{k}=\left(\frac{\hbar\mathbf{k}^2}{2m}-\mu\right)\hat{\tau}_z+i\left(\widetilde{\Delta}\hat{\tau}_+-\widetilde{\Delta}^*\hat{\tau}_-\right)\otimes\hat{s}_y.
\end{equation}
Solving the corresponding BdG eigenvalue equation yields the longitudinal wave vectors of the electronlike and holelike quasiparticles as
\begin{equation}
	k_x^\pm=\sqrt{\bar{k}_x^2\pm\frac{2m}{\hbar^2}\sqrt{E^2-\Delta_T^2}},
\end{equation}
where $\bar{k}=\sqrt{k_F^2-k_y^2}$ and $k_y$ is the transverse wave vector parallel to the interface.
Considering the electronlike and holelike quasiparticles with both spin orientations and both propagation directions along the $x$ axis, we obtain eight independent quasiparticle eigenmodes in the left superconducting region $x<0$, as listed in Table~\ref{LS_eigen_state}.
Here, we define
\begin{equation}
	\mathbb{L}_{e\uparrow}=
	\begin{bmatrix}
		e^{i\phi/2} \\ 0 \\ 0 \\ \gamma e^{-i\phi/2}
	\end{bmatrix},~
	\mathbb{L}_{e\downarrow}=
	\begin{bmatrix}
		0\\ e^{i\phi/2} \\ -\gamma e^{-i\phi/2} \\ 0
	\end{bmatrix},~
	\mathbb{L}_{h\downarrow}=
	\begin{bmatrix}
		\gamma e^{i\phi/2} \\ 0 \\ 0 \\ e^{-i\phi/2}
	\end{bmatrix},~
	\mathbb{L}_{h\uparrow}=
	\begin{bmatrix}
		0 \\ \gamma e^{i\phi/2} \\ -e^{-i\phi/2} \\ 0
	\end{bmatrix},
\end{equation}
where the coherence factor is given by $\gamma=\Delta_T/(E+\sqrt{E^2-\Delta_T^2})$.

\begin{table*}
	\caption{\label{LS_eigen_state}
		Quasiparticle eigenmodes in the left superconducting region.
	}
	\begin{ruledtabular}
		\begin{tabular}{lcccc}
			\textrm{Direction}&
			\textrm{$e\uparrow$}&
			\textrm{$e\downarrow$}&
			\textrm{$h\downarrow$}&
			\textrm{$h\uparrow$}\\
			\colrule
			Right & $\mathbb{L}_{e\uparrow}e^{ik_x^+x}e^{ik_y y}$ & $\mathbb{L}_{e\downarrow}e^{ik_x^+x}e^{ik_y y}$ & $\mathbb{L}_{h\downarrow}e^{-ik_x^-x}e^{ik_y y}$ & $\mathbb{L}_{h\uparrow}e^{-ik_x^-x}e^{ik_y y}$ \\
			Left & $\mathbb{L}_{e\uparrow}e^{-ik_x^+x}e^{ik_y y}$ & $\mathbb{L}_{e\downarrow}e^{-ik_x^+x}e^{ik_y y}$ & $\mathbb{L}_{h\downarrow}e^{ik_x^-x}e^{ik_y y}$ & $\mathbb{L}_{h\uparrow}e^{ik_x^-x}e^{ik_y y}$ \\
		\end{tabular}
	\end{ruledtabular}
\end{table*}

We next consider the AM region \(0<x<L\).
In this region, $\widetilde{\Delta}=0$ and $\widetilde{\mu}=\mu+\delta\mu$.
The BdG Hamiltonian then reduces to
\begin{equation}
	\hat{\mathcal{H}}_\mathbf{k}=\left(\frac{\hbar\mathbf{k}^2}{2m}-\widetilde{\mu}\right)\hat{\tau}_z+M_\mathbf{k}\hat{\tau}_z\otimes\hat{s}_z.
\end{equation}
For convenience, we rewrite the altermagnetic term $M_\mathbf{k}$ as
\begin{equation}
	M_\mathbf{k}=\alpha_1 k_x k_y + \frac{1}{2}\alpha_2\left(k_x^2-k_y^2\right),
\end{equation}
where $\alpha_1=2J\sin(2\alpha)/k_F^2$ and $\alpha_2=2J\cos(2\alpha)/k_F^2$.
Notably, the Hamiltonian is already diagonal, and its four eigenvalues and corresponding eigenvectors are
\begin{equation}
	E=\xi_\mathbf{k}+M_\mathbf{k},~\psi_{e\uparrow}=
	\begin{bmatrix}
		1 \\ 0\\ 0 \\ 0
	\end{bmatrix};~
	E=\xi_\mathbf{k}-M_\mathbf{k},~\psi_{e\downarrow}=
	\begin{bmatrix}
		0 \\ 1 \\ 0 \\ 0
	\end{bmatrix};~
	E=-\xi_\mathbf{k}-M_\mathbf{k},~\psi_{h\uparrow}=
	\begin{bmatrix}
		0 \\ 0 \\ 1 \\ 0
	\end{bmatrix};~
	E=-\xi_\mathbf{k}+M_\mathbf{k},~\psi_{h\downarrow}=
	\begin{bmatrix}
		0\\0\\0\\1
	\end{bmatrix}.
\end{equation}
Solving the corresponding dispersion relations yields the $x$ components of the wave vectors for the electron and hole quasiparticles
\begin{equation}
	k_{e\uparrow}^\pm=\frac{-m\alpha_1 k_y\pm \mathcal{D}_+(E)}{\hbar^2+m\alpha_2},~
	k_{e\downarrow}^\pm=\frac{m\alpha_1 k_y\pm \mathcal{D}_-(E)}{\hbar^2-m\alpha_2},~
	k_{h\uparrow}^\pm=\frac{-m\alpha_1 k_y\pm \mathcal{D}_+(-E)}{\hbar^2+m\alpha_2},~
	k_{h\downarrow}^\pm=\frac{m\alpha_1 k_y\pm \mathcal{D}_-(-E)}{\hbar^2-m\alpha_2}.
	\label{AM_Wave_vector}
\end{equation}
Here, we have introduced the auxiliary quantity
\begin{equation}
	\mathcal{D}_\pm(E)=\hbar\sqrt{2m\left(1\pm\frac{m\alpha_2}{\hbar^2}\right)\left(\mu+\delta\mu+E\right)-\hbar^2k_y^2+\frac{mk_y^2\left(\alpha_1^2+\alpha_2^2\right)}{\hbar^2}}.
\end{equation}
Combining the four eigenvectors with the right- and left-going wave-vector solutions yields eight independent quasiparticle eigenmodes in the AM region, as listed in Table~\ref{AM_eigen_state}.
Here, we have defined
\begin{equation}
	\mathbb{M}_{e\uparrow}=\psi_{e\uparrow},~\mathbb{M}_{e\downarrow}=\psi_{e\downarrow},~
	\mathbb{M}_{h\uparrow}=\psi_{h\uparrow},~\mathbb{M}_{h\downarrow}=\psi_{h\downarrow}.
\end{equation}

\begin{table*}
	\caption{\label{AM_eigen_state}
		Quasiparticle eigenmodes in the AM region.
	}
	\begin{ruledtabular}
		\begin{tabular}{lcccc}
			\textrm{Direction}&
			\textrm{$e\uparrow$}&
			\textrm{$e\downarrow$}&
			\textrm{$h\uparrow$}&
			\textrm{$h\downarrow$}\\
			\colrule
			Right & $\mathbb{M}_{e\uparrow}e^{ik_{e\uparrow}^+x}e^{ik_y y}$ & $\mathbb{M}_{e\downarrow}e^{ik_{e\downarrow}^+x}e^{ik_y y}$ & $\mathbb{M}_{h\uparrow}e^{ik_{h\uparrow}^-x}e^{ik_y y}$ & $\mathbb{M}_{h\downarrow}e^{ik_{h\downarrow}^-x}e^{ik_y y}$ \\
			Left & $\mathbb{M}_{e\uparrow}e^{ik_{e\uparrow}^-x}e^{ik_y y}$ & $\mathbb{M}_{e\downarrow}e^{ik_{e\downarrow}^-x}e^{ik_y y}$ & $\mathbb{M}_{h\uparrow}e^{ik_{h\uparrow}^+x}e^{ik_y y}$ & $\mathbb{M}_{h\downarrow}e^{ik_{h\downarrow}^+x}e^{ik_y y}$ \\
		\end{tabular}
	\end{ruledtabular}
\end{table*}

Applying the same procedure to the right superconducting region $x>L$ yields eight independent quasiparticle eigenmodes, as listed in Table~\ref{RS_eigen_state}.
Here, we define
\begin{equation}
	\mathbb{R}_{e\uparrow}=
	\begin{bmatrix}
		1 \\ 0 \\ 0 \\ \gamma
	\end{bmatrix},~
	\mathbb{R}_{e\downarrow}=
	\begin{bmatrix}
		0 \\ 1 \\ -\gamma \\ 0
	\end{bmatrix},~
	\mathbb{R}_{h\downarrow}=
	\begin{bmatrix}
		\gamma \\ 0 \\ 0 \\ 1
	\end{bmatrix},~
	\mathbb{R}_{h\uparrow}=
	\begin{bmatrix}
		0 \\ \gamma \\ -1 \\ 0
	\end{bmatrix}.
\end{equation}

\begin{table*}
	\caption{\label{RS_eigen_state}
		Quasiparticle eigenmodes in the right superconducting region.
	}
	\begin{ruledtabular}
		\begin{tabular}{lcccc}
			\textrm{Direction}&
			\textrm{$e\uparrow$}&
			\textrm{$e\downarrow$}&
			\textrm{$h\downarrow$}&
			\textrm{$h\uparrow$}\\
			\colrule
			Right & $\mathbb{R}_{e\uparrow}e^{ik_x^+x}e^{ik_y y}$ & $\mathbb{R}_{e\downarrow}e^{ik_x^+x}e^{ik_y y}$ & $\mathbb{R}_{h\downarrow}e^{-ik_x^-x}e^{ik_y y}$ & $\mathbb{R}_{h\uparrow}e^{-ik_x^-x}e^{ik_y y}$ \\
			Left & $\mathbb{R}_{e\uparrow}e^{-ik_x^+x}e^{ik_y y}$ & $\mathbb{R}_{e\downarrow}e^{-ik_x^+x}e^{ik_y y}$ & $\mathbb{R}_{h\downarrow}e^{ik_x^-x}e^{ik_y y}$ & $\mathbb{R}_{h\uparrow}e^{ik_x^-x}e^{ik_y y}$ \\
		\end{tabular}
	\end{ruledtabular}
\end{table*}

\subsection{\label{ABSS}Andreev Bound-State Spectrum}
\subsubsection{Derivation of the Andreev Energy Spectrum}
In Sec.~\ref{ES_BdG_Eq}, we derived the quasiparticle eigenmodes and their corresponding wave vectors in each region of the Josephson junction.
We consider the short-junction limit ($L\ll\xi_0$), where $\xi_0=\hbar v_F/\Delta_0$ is the coherence length.
In this limit, the Josephson current is dominated by the discrete Andreev bound states.
We therefore analyze how the junction parameters affect the Andreev bound-state spectrum and the resulting Josephson transport.
The $x$ components of the wave vectors for the electron and hole quasiparticles in the AM region are given by Eq.~\eqref{AM_Wave_vector}.
Within the Andreev approximation, we evaluate the quasiparticle wave vectors in the AM region at the Fermi energy.
Equation~\eqref{AM_Wave_vector} then reduces to
\begin{equation}
	k_{e\uparrow}^\pm=\frac{-m\alpha_1 k_y\pm \mathcal{D}_+}{\hbar^2+m\alpha_2},~
	k_{e\downarrow}^\pm=\frac{m\alpha_1 k_y\pm \mathcal{D}_-}{\hbar^2-m\alpha_2},~
	k_{h\uparrow}^\pm=\frac{-m\alpha_1 k_y\pm \mathcal{D}_+}{\hbar^2+m\alpha_2},~
	k_{h\downarrow}^\pm=\frac{m\alpha_1 k_y\pm \mathcal{D}_-}{\hbar^2-m\alpha_2},
	\label{AM_Wave_vector_FS}
\end{equation}
where
\begin{equation}
	\mathcal{D}_\pm=\hbar\sqrt{2m\left(1\pm\frac{m\alpha_2}{\hbar^2}\right)\left(\mu+\delta\mu\right)-\hbar^2k_y^2+\frac{mk_y^2\left(\alpha_1^2+\alpha_2^2\right)}{\hbar^2}}.
\end{equation}
Substituting $\mathcal{D}_z=\pm$ into Eq.~\eqref{AM_Wave_vector_FS} and rearranging the terms, we obtain
\begin{subequations}
	\begin{align}
		k_{e\uparrow}^{\mathcal{D}_z}&=\frac{-m\alpha_1 k_y\left(\hbar^2-m\alpha_2\right)+\mathcal{D}_z\mathcal{D}_+\left(\hbar^2-m\alpha_2\right)}{\hbar^4-m^2\alpha_2^2},\\
		k_{e\downarrow}^{\mathcal{D}_z}&=\frac{m\alpha_1 k_y\left(\hbar^2+m\alpha_2\right)+\mathcal{D}_z\mathcal{D}_-\left(\hbar^2+m\alpha_2\right)}{\hbar^4-m^2\alpha_2^2},\\
		k_{h\uparrow}^{\mathcal{D}_z}&=\frac{-m\alpha_1 k_y\left(\hbar^2-m\alpha_2\right)+\mathcal{D}_z\mathcal{D}_+\left(\hbar^2-m\alpha_2\right)}{\hbar^4-m^2\alpha_2^2},\\
		k_{h\downarrow}^{\mathcal{D}_z}&=\frac{m\alpha_1 k_y\left(\hbar^2+m\alpha_2\right)+\mathcal{D}_z\mathcal{D}_-\left(\hbar^2+m\alpha_2\right)}{\hbar^4-m^2\alpha_2^2}.
	\end{align}
\end{subequations}
To clarify the physical meaning of the above result, we rewrite $\mathcal{D}_\pm=\bar{\mathcal{D}}\pm\delta D$.
In terms of $\bar{\mathcal{D}}$ and $\delta D$, the $x$ components of the electron and hole quasiparticle wave vectors can be written as
\begin{equation}
	\mathcal{K}_{e\uparrow}^\pm=\mathcal{K}_{h\uparrow}^\pm=\mathcal{K}_0\pm+\mathcal{K}_z^\pm,~
	\mathcal{K}_{e\downarrow}^\pm=\mathcal{K}_{h\downarrow}^\pm=\mathcal{K}_0\pm-\mathcal{K}_z^\pm,
\end{equation}
where
\begin{subequations}
	\begin{align}
		\mathcal{K}_0^\pm&=\frac{\pm\left(\hbar^2\bar{\mathcal{D}}-m\alpha_2\delta D\right)+m^2\alpha_1\alpha_2 k_y}{\hbar^4-m^2\alpha_2^2},\\
		\mathcal{K}_z^\pm&=\frac{\pm\left(\hbar^2\delta D-m\alpha_2\bar{\mathcal{D}}\right)-\hbar^2 m\alpha_1 k_y}{\hbar^4-m^2\alpha_2^2}.
	\end{align}
\end{subequations}

%%%%%%%%%%%%%%%%%%%%%%%%%%%%%%%%%%%%%%%%%%%%%%%%%%%%%%%%%%%%%%%%%%%%%%%%%%%%%%
A bound state forms when the total phase accumulated over a complete electron-hole round trip between the two interfaces is an integer multiple of $2\pi$.
We first consider perfectly transparent interfaces, for which normal reflection is absent.
At a superconducting interface with pairing potential $\Delta_Te^{i\phi}$, an electron of energy $E$ is Andreev reflected as a hole and acquires the phase factor $\exp[i(-\phi-\alpha(E))]$~\cite{1991BeenakkerP38363839,2023BeenakkerP7542575425,2021AsanoP}.
Conversely, a hole Andreev reflected as an electron acquires $\exp[i(\phi-\alpha(E))]$~\cite{1991BeenakkerP38363839,2023BeenakkerP7542575425,2021AsanoP}.
Here, $\alpha(E)=\arccos(E/\Delta_T)\in (0,\pi)$ and $\vert E\vert<\Delta_T$.
In the short-junction limit, the energy dependence of the propagation phase accumulated by the quasiparticles in the AM region can be neglected, whereas that of the Andreev reflection phase must be retained.
The resulting Andreev bound-state energies are
\begin{subequations}
	\begin{align}
		E_{e\uparrow}=\Delta_T\cos\left(\frac{\phi+2\mathcal{K}_z^+L}{2}\right)\cos(n\pi),\\
		E_{e\downarrow}=\Delta_T\cos\left(\frac{\phi-2\mathcal{K}_z^+L}{2}\right)\cos(n\pi),\\
		E_{h\uparrow}=\Delta_T\cos\left(\frac{\phi-2\mathcal{K}_z^-L}{2}\right)\cos(n\pi),\\
		E_{h\downarrow}=\Delta_T\cos\left(\frac{\phi+2\mathcal{K}_z^-L}{2}\right)\cos(n\pi).
	\end{align}
	\label{Andreev_Level_Condition}
\end{subequations}
Equation~\eqref{Andreev_Level_Condition} shows that the phase contributions from $\mathcal{K}_0^\pm$ cancel exactly over a complete electron-hole round trip, leaving only the contributions from $\mathcal{K}_z^\pm$ to the net accumulated phase.

%%%%%%%%%%%%%%%%%%%%%%%%%%%%%%%%%%%%%%%%%%%%%%%%%%%%%%%%%%%%%%%%%%%%%%%%%%%%%%
In the preceding analysis, we assumed perfectly transparent interfaces, for which normal reflection is absent, to elucidate the underlying physical mechanism.
We now turn to the case with normal reflection.
To obtain an analytical result, we further simplify the model.
Specifically, we consider a $d_{xy}$-wave altermagnet, for which the Hamiltonian reduces to
\begin{equation}
	\hat{\mathcal{H}}_\mathbf{k}=\left(\frac{\hbar\mathbf{k}^2}{2m}-\widetilde{\mu}+U\right)\hat{\tau}_z+\alpha_1k_xk_y\hat{\tau}_z\otimes\hat{s}_z+i\left(\widetilde{\Delta}\hat{\tau}_+-\widetilde{\Delta}^*\hat{\tau}_-\right)\otimes\hat{s}_y.
	\label{H_S_AMXY_S}
\end{equation}
We then change the Nambu basis from
$[\hat{c}_{\mathbf{k},\uparrow},\hat{c}_{\mathbf{k},\downarrow},\hat{c}_{-\mathbf{k},\uparrow}^\dagger,\hat{c}_{-\mathbf{k},\downarrow}^\dagger]$
to
$[\hat{c}_{\mathbf{k},\uparrow},\hat{c}_{\mathbf{k},\downarrow},\hat{c}_{-\mathbf{k},\downarrow}^\dagger,-\hat{c}_{-\mathbf{k},\uparrow}^\dagger]$.
In this new basis, the Hamiltonian in Eq.~\eqref{H_S_AMXY_S} can be written as
\begin{equation}
	\hat{\mathcal{H}}_\mathbf{k}=\left(\frac{\hbar\mathbf{k}^2}{2m}-\widetilde{\mu}(x)+U(x)\right)\hat{\tau}_z+\frac{1}{2}\frac{\hbar^2}{m}\left[\widetilde{\alpha}_1(x)k_x+k_x\widetilde{\alpha}_1(x)\right]k_y\hat{s}_z+\Delta_T(x)\left[\hat{\tau}_x\cos\phi(x)-\hat{\tau}_y\sin\phi(x)\right],
	\label{H_S_AMXY_S_New_basis}
\end{equation}
where $\widetilde{\alpha}_1=m\alpha_1/\hbar^2$.
The spatial dependence of the parameters specifies the junction profile, and the anticommutator $\{\widetilde{\alpha}_1(x),k_x\}$ ensures the Hermiticity of the Hamiltonian~\eqref{H_S_AMXY_S_New_basis}.
Applying the unitary transformation $\hat{U}=\exp[i\hat{\tau}_z\hat{s}_zk_y\int_0^x\widetilde{\alpha}_1(x^\prime)dx^\prime]$, we obtain the transformed Hamiltonian
\begin{equation}
	\hat{U}\hat{\mathcal{H}}_\mathbf{k}\hat{U}^\dagger=\left(\frac{1}{2}\frac{\hbar^2}{m}k_x^2+\frac{1}{2}\frac{\hbar^2}{m}\left(1-\widetilde{\alpha}_1k_y^2\right)-\widetilde{\mu}(x)+U(x)\right)\hat{\tau}_z+\Delta_T(x)\left[\hat{\tau}_x\cos\widetilde{\phi}(x)-\hat{\tau}_y\sin\widetilde{\phi}(x)\right],
	\label{H_S_AMXY_S_New_basis_UT}
\end{equation}
where
$\widetilde{\phi}(x)=\phi(x)+2\hat{s}_z\int_0^x\widetilde{\alpha}_1(x^\prime)dx^\prime=\phi(x)\pm 2 k_y \widetilde{\alpha}_1L$
denotes the spin-dependent phase difference.
Therefore, for a fixed $k_y$ and spin, the AMJJ is equivalent to a nonmagnetic Josephson junction with the effective phase difference $\widetilde{\phi}(x)$.
Using the known result for the Andreev bound-state spectrum of a nonmagnetic Josephson junction, we obtain the spectrum of the AMJJ~\cite{2023BeenakkerP7542575425,1991BeenakkerP38363839,1999FurusakiP809818}
\begin{subequations}
	\begin{align}
		E_\uparrow^\pm=\pm\Delta_T\sqrt{1-T\left(k_y\right)\sin^2\left(\frac{\phi}{2}-k_y\widetilde{\alpha}_1L\right)},\\
		E_\downarrow^\pm=\pm\Delta_T\sqrt{1-T\left(k_y\right)\sin^2\left(\frac{\phi}{2}+k_y\widetilde{\alpha}_1L\right)}.
	\end{align}
	\label{Andreev_Level_Reflect}
\end{subequations}
Here, $T(k_y)$ denotes the transmission probability through the junction evaluated in the normal state ($\Delta_T=0$)
\begin{equation}
	T\left(k_y\right)=\frac{\Gamma^2}{2\left(1-\Gamma\right)\cos\left(2\mathcal{K}_0^+\right)+1+\left(1-\Gamma\right)^2},
\end{equation}
where $\mathcal{K}_0^+=-\mathcal{K}_0^-=1/\hbar\sqrt{2m(\mu+\delta\mu)-\hbar^2k_y^2+m^2k_y^2\alpha_1^2/\hbar^2}$.
Here, $\Gamma$ denotes the normal-state interfacial tunnel probability.
To evaluate $\Gamma$, we set $\Delta_0=0$ and consider a $\delta$-function barrier $U(x)=U_I\delta(x)$ at $x=0$.
Following the standard BTK scattering formalism~\cite{1982BlonderP45154532},
the scattering states are written as $\psi_S=e^{ik_xx}+re^{-ik_xx}$ and $\psi_{\mathrm{AM}}=te^{ik_{\mathrm{AM}x}x}$, where $r$ and $t$ are the reflection and transmission amplitudes, respectively.
Continuity of the wave function and the derivative discontinuity induced by the $\delta$-function barrier give the boundary conditions
\begin{subequations}
	\begin{align}
		1+r &= t,\\
		\left.\partial_x\psi\right|_{x=0^+}
		-\left.\partial_x\psi\right|_{x=0^-}
		&=\frac{2mU_I}{\hbar^2}\psi(0).
	\end{align}
\end{subequations}
Solving these boundary conditions gives
\begin{equation}
	t=\frac{2ik_x}{i(k_{\mathrm{AM}x}+k_x)-2Z}.
\end{equation}
The transmission probability is determined from the ratio of the transmitted and incident probability currents
\begin{equation}
	\Gamma(k_y)
	=\frac{k_{\mathrm{AM}x}}{k_x}|t|^2
	=\frac{4k_xk_{\mathrm{AM}x}}
	{(k_x+k_{\mathrm{AM}x})^2+4Z^2}.
\end{equation}
The longitudinal wave-vector components of the quasiparticles in the superconducting and AM regions are given by
\begin{equation}
	k_x=\sqrt{\frac{2m\mu}{\hbar^2 k_F^2}-k_y^2},~k_{\mathrm{AM}x}=\sqrt{\frac{2m\left(\mu+\delta\mu\right)}{\hbar^2k_F^2}-\left(1-\widetilde{\alpha}_1^2\right)k_y^2}.
\end{equation}
We emphasize that both $k_x$ and $k_{\mathrm{AM}x}$ are obtained from the Hamiltonian~\eqref{H_S_AMXY_S_New_basis_UT}.

%%%%%%%%%%%%%%%%%%%%%%%%%%%%%%%%%%%%%%%%%%%%%%%%%%%%%%%%%%%%%%%%%%%%%%%%%%%%%
%In the short-junction limit, the Josephson current is dominated by the Andreev bound states, while the contribution from the continuum states can be neglected.
%At thermal equilibrium at temperature $T$, the Josephson free energy of the junction is given by
%\begin{equation}
%	\mathcal{F}_\mathrm{Ana}=-\sum_{E>0}\frac{1}{2}E\tanh\left(\frac{1}{2}\beta E\right),
%	\label{F_Ana}
%\end{equation}
%where $\beta=1/(k_B T)$ is the inverse temperature, and $\sum_{E>0}$ runs over all positive-energy Andreev bound states satisfying $(0,\Delta_T)$, including all transverse momenta $k_y$ and both spin states $\uparrow$ and $\downarrow$.
%Equation \eqref{F_Ana} shows that the Josephson free energy is obtained by summing the contributions of the Andreev bound states from different transverse-momentum channels $k_y$.
%Thus, introducing an altermagnet enables control over the phase shifts of different transverse-momentum channels, thereby modifying how their contributions combine and potentially giving rise to unconventional Josephson effects.

%%%%%%%%%%%%%%%%%%%%%%%%%%%%%%%%%%%%%%%%%%%%%%%%%%%%%%%%%%%%%%%%%%%%%%%%%%%%%
\subsubsection{Numerical Analysis of the Andreev Energy Spectrum}
\begin{figure}
	\centering
	\includegraphics[width=1\textwidth]{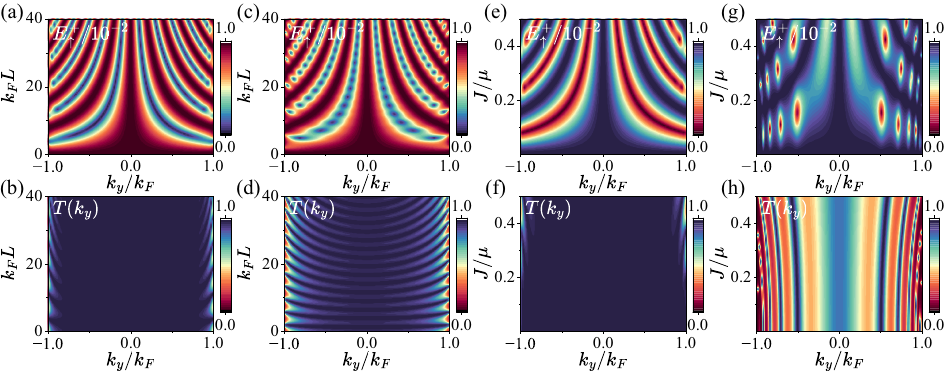}
	\caption{Dependence of the Andreev bound-state spectrum $E_{\uparrow}^{+}$ and the transmission probability $T(k_y)$ on the thickness $L$ of the altermagnetic layer and the exchange interaction strength $J$ for different transverse-momentum channels. The parameters are listed in Table~\ref{Para_SMFig1}.}
	\label{SMFig1}
\end{figure}

Here, we focus on a Josephson junction incorporating a pure $d_{xy}$-wave altermagnet, for which the Andreev spectrum can be obtained analytically.
Equation \eqref{Andreev_Level_Reflect} shows that each transverse-momentum channel behaves as a Josephson channel with a $k_y$-dependent phase shift, while the macroscopic Josephson free energy is obtained by summing the contributions from all channels.
Furthermore, the transmission probability $T(k_y)$ in the normal state ($\Delta_0=0$) is governed by the S-AM-S interfacial barrier parameter $Z$ and the potential offset $\delta\mu$.
To gain physical insight into the junction, we first analyze how these microscopic parameters independently modulate the Andreev spectrum.
Unless stated otherwise, we use the temperature-dependent superconducting gap $\Delta(T)=\Delta_0\tanh(1.74\sqrt{T/T_c-1})$, where $T_c$ is the superconducting critical temperature and $\Delta_0=1.76k_BT_c$ is the zero-temperature gap.
In the numerical calculations, we set $\Delta_0=0.01\mu$.
Figure \ref{SMFig1} shows the Andreev spectrum at the superconducting phase difference $\phi=0$.
At $\phi=0$,, the two spin-resolved branches are degenerate, i.e., $E_\uparrow^\pm=E_\downarrow^\pm$.

%%%%%%%%%%%%%%%%%%%%%%%%%%%%%%%%%%%%%%%%%%%%%%%%%%%%%%%%%%%%%%%%%%%%%%%%%%%%%%%%%
As shown in Fig.~\ref{SMFig1}(a), increasing the junction length $L$ causes the Andreev levels to oscillate more rapidly with the transverse momentum $k_y$.
Figure \ref{SMFig1}(b) shows the corresponding normal-state transmission probability $T(k_y)$.
As shown in Figs.~\ref{SMFig1}(c) and \ref{SMFig1}(d), introducing $\delta\mu$ produces a momentum-selective filtering effect through $T(k_y)$, thereby masking the Andreev spectrum.
As shown in Fig.~\ref{SMFig1}(e), increasing the exchange interaction strength $J$ leads to a stronger $k_y$ dependence of the phase shifts of the Andreev levels in different momentum channels.
Figure \ref{SMFig1}(f) further shows that $J$ appreciably affects only the channels with $k_y$ close to $k_F$, while having little effect on the other channels.
We next examine the effect of the interface-barrier parameter $Z$.
As shown in Fig.~\ref{SMFig1}(g), introducing $Z$ leaves the overall oscillatory fringe pattern of the Andreev spectrum essentially unchanged, while the modifications to the spectrum arise mainly from the corresponding changes in the transmission probability $T(k_y)$.
Figure \ref{SMFig1}(h) further shows that $Z$ substantially modifies $T(k_y)$, which acts as a mask to select the contributing Andreev bound states.
In this picture, $J$ and $L$ tune the channel-dependent phase shifts, whereas $Z$ and $\delta\mu$ regulate the relative channel contributions.

\begin{table*}
	\caption{\label{Para_SMFig1}
		Parameters used in Fig.~\ref{SMFig1}.
	}
	\begin{ruledtabular}
		\begin{tabular}{lccccc}
			\textrm{No.}&
			\textrm{$\alpha$}&
			\textrm{$J/\mu$}&
			\textrm{$k_FL$}&
			\textrm{$\delta\mu/\mu$}&
			\textrm{$Z$}\\
			\colrule
			Fig.~\ref{SMFig1}(a,~b) & $\pi/4$ & $0.4$ & $0\sim 40$ & $0$ & $0$ \\
			Fig.~\ref{SMFig1}(c,~d) & $\pi/4$ & $0.4$ & $0\sim 40$ & $0.5$ & $0$ \\
			Fig.~\ref{SMFig1}(e,~f) & $\pi/4$ & $0\sim0.5$ & $20$ & $0$ & $0$ \\
			Fig.~\ref{SMFig1}(g,~h) & $\pi/4$ & $0\sim0.5$ & $20$ & $0$ & $0.5$ \\
		\end{tabular}
	\end{ruledtabular}
\end{table*}

\subsection{Full Numerical Solution}
In Sec.~\ref{ABSS}, we employed a semianalytical approach to investigate the effects of different parameters on the properties of the Josephson junction from the perspective of the Andreev bound states.
The results show that the altermagnet parameters primarily control the phase shifts of individual momentum channels, thereby modifying how the contributions from different channels combine.
In contrast, the interface parameters of the Josephson junction filter the Andreev bound states in individual momentum channels that contribute to the Josephson current.
The combined effects of these two sets of parameters enable the Josephson energy to be engineered on demand.
To characterize the effects of these parameters on the Josephson junction more accurately, we now turn to a fully numerical analysis.

%%%%%%%%%%%%%%%%%%%%%%%%%%%%%%%%%%%%%%%%%%%%%%%%%%%%%%%%%%%%%%%%%%%%%%%%%%%%%
In Sec.~\ref{ES_BdG_Eq}, we obtained the eigenfunctions in each region of the Josephson junction.
The quasiparticle wave functions in the three regions can then be constructed from these eigenfunctions.
In the left superconducting region, the quasiparticle wave function is given by
\begin{subequations}
	\begin{align}
		\psi_{e\uparrow}^L&=\left(\mathbb{L}_{e\uparrow}e^{ik_x^+x}+a_{e\uparrow}\mathbb{L}_{h\downarrow}e^{ik_x^-x}+r_{e\uparrow}\mathbb{L}_{e\uparrow}e^{-ik_x^+x}\right)e^{ik_yy},\\
		\psi_{h\downarrow}^L&=\left(\mathbb{L}_{h\downarrow}e^{-ik_x^-x}+a_{h\downarrow}\mathbb{L}_{e\uparrow}e^{-ik_x^+x}+r_{h\downarrow}\mathbb{L}_{h\downarrow}e^{ik_x^-x}\right)e^{ik_yy},\\
		\psi_{e\downarrow}^L&=\left(\mathbb{L}_{e\downarrow}e^{ik_x^+x}+a_{e\downarrow}\mathbb{L}_{h\uparrow}e^{ik_x^-x}+r_{e\downarrow}\mathbb{L}_{e\downarrow}e^{-ik_x^+x}\right)e^{ik_yy},\\
		\psi_{h\uparrow}^L&=\left(\mathbb{L}_{h\uparrow}e^{-ik_x^-x}+a_{h\uparrow}\mathbb{L}_{e\downarrow}e^{-ik_x^+x}+r_{h\uparrow}\mathbb{L}_{h\uparrow}e^{ik_x^-x}\right)e^{ik_yy}.
	\end{align}
\end{subequations}
In the altermagnetic region, the quasiparticle wave function takes the form
\begin{subequations}
	\begin{align}
		\psi_{e\uparrow}^{M}
		&=
		\left(
		s_{11}\mathbb{M}_{e\uparrow}e^{ik_{e\uparrow}^{+}x}
		+s_{12}\mathbb{M}_{e\uparrow}e^{ik_{e\uparrow}^{-}x}
		+s_{13}\mathbb{M}_{h\downarrow}e^{ik_{h\downarrow}^{+}x}
		+s_{14}\mathbb{M}_{h\downarrow}e^{ik_{h\downarrow}^{-}x}
		\right)e^{ik_y y},
		\\
		\psi_{h\downarrow}^{M}
		&=
		\left(
		s_{21}\mathbb{M}_{e\uparrow}e^{ik_{e\uparrow}^{+}x}
		+s_{22}\mathbb{M}_{e\uparrow}e^{ik_{e\uparrow}^{-}x}
		+s_{23}\mathbb{M}_{h\downarrow}e^{ik_{h\downarrow}^{+}x}
		+s_{24}\mathbb{M}_{h\downarrow}e^{ik_{h\downarrow}^{-}x}
		\right)e^{ik_y y},
		\\
		\psi_{e\downarrow}^{M}
		&=
		\left(
		s_{31}\mathbb{M}_{e\downarrow}e^{ik_{e\downarrow}^{+}x}
		+s_{32}\mathbb{M}_{e\downarrow}e^{ik_{e\downarrow}^{-}x}
		+s_{33}\mathbb{M}_{h\uparrow}e^{ik_{h\uparrow}^{+}x}
		+s_{34}\mathbb{M}_{h\uparrow}e^{ik_{h\uparrow}^{-}x}
		\right)e^{ik_y y},
		\\
		\psi_{h\uparrow}^{M}
		&=
		\left(
		s_{41}\mathbb{M}_{e\downarrow}e^{ik_{e\downarrow}^{+}x}
		+s_{42}\mathbb{M}_{e\downarrow}e^{ik_{e\downarrow}^{-}x}
		+s_{43}\mathbb{M}_{h\uparrow}e^{ik_{h\uparrow}^{+}x}
		+s_{44}\mathbb{M}_{h\uparrow}e^{ik_{h\uparrow}^{-}x}
		\right)e^{ik_y y}.
	\end{align}
\end{subequations}
In the right superconducting region, the quasiparticle wave function is given by
\begin{subequations}
	\begin{align}
		\psi_{e\uparrow}^{R}
		&=
		\left(
		c_{1}\mathbb{R}_{e\uparrow}e^{ik_x^{+}x}
		+d_{1}\mathbb{R}_{h\downarrow}e^{-ik_x^{-}x}
		\right)e^{ik_y y},
		\\
		\psi_{h\downarrow}^{R}
		&=
		\left(
		c_{2}\mathbb{R}_{e\uparrow}e^{ik_x^{+}x}
		+d_{2}\mathbb{R}_{h\downarrow}e^{-ik_x^{-}x}
		\right)e^{ik_y y},
		\\
		\psi_{e\downarrow}^{R}
		&=
		\left(
		c_{3}\mathbb{R}_{e\downarrow}e^{ik_x^{+}x}
		+d_{3}\mathbb{R}_{h\uparrow}e^{-ik_x^{-}x}
		\right)e^{ik_y y},
		\\
		\psi_{h\uparrow}^{R}
		&=
		\left(
		c_{4}\mathbb{R}_{e\downarrow}e^{ik_x^{+}x}
		+d_{4}\mathbb{R}_{h\uparrow}e^{-ik_x^{-}x}
		\right)e^{ik_y y}.
	\end{align}
\end{subequations}
The wave functions in the three regions are matched by imposing the following boundary conditions at the two interfaces ($x=0$ and $x=L$):
\begin{subequations}
	\begin{align}
		\left.\psi\right|_{x=0^{-}}&=\left.\psi\right|_{x=0^{+}},\\
		\left(\frac{\hbar^{2}}{m}+\alpha_{2}\hat{s}_{z}\right)
		\left.\left(-i\partial_{x}\psi\right)\right|_{x=0^{+}}
		-\frac{\hbar^{2}}{m}\left.\left(-i\partial_{x}\psi\right)\right|_{x=0^{-}}
		&=\left(-\alpha_{1}k_{y}\hat{s}_{z}-2iU_{I}\right)\left.\psi\right|_{x=0},\\
		\left.\psi\right|_{x=L^{-}}&=\left.\psi\right|_{x=L^{+}},\\
		\frac{\hbar^{2}}{m}\left.\left(-i\partial_{x}\psi\right)\right|_{x=L^{+}}
		-\left(\frac{\hbar^{2}}{m}+\alpha_{2}\hat{s}_{z}\right)\left.\left(-i\partial_{x}\psi\right)\right|_{x=L^{-}}
		&=\left(i\alpha_{1}k_{y}\hat{s}_{z}-2iU_{I}\right)\left.\psi\right|_{x=L}.
	\end{align}
\end{subequations}
The Josephson current arises from four Andreev reflection processes, whose reflection coefficients are denoted by $a_{e\uparrow}$, $a_{h\uparrow}$, $a_{e\downarrow}$, and $a_{h\downarrow}$.
Using the Furusaki-Tsukada formula, the Josephson current is given by
\begin{equation}
	I(\phi)=\int_{-\pi/2}^{\pi/2}I(\theta)d\theta,
\end{equation}
where $I(\theta)$ denotes the contribution of each transverse-momentum channel to the Josephson current and is defined by
\begin{equation}
	I(\theta)=\frac{e\Delta_T\cos\theta}{2\hbar\beta}\sum_{\omega_n,s}\frac{k_{nx}^++k_{nx}^-}{\sqrt{\omega_n^2+\Delta_T^2}}\left(\frac{a_{e,s}}{k_{nx}^+}-\frac{a_{h,s}}{k_{nx}^-}\right).
\end{equation}
Here, we take the transverse momentum to be $k_y=k_F\sin\theta$ and apply the analytic continuation $E\rightarrow i\omega_n$ to the incident quasiparticle energy.
Here, $\omega_n=\pi k_BT(2n+1)$ denotes the $n$th Matsubara frequency, with $n=0,\pm1,\pm2,\dots$.
We normalize the Josephson current using the factor $2eR_NI/\pi\Delta_0$, where $R_N$ denotes the normal-state resistance of the normal-metal-altermagnet-normal-metal junction.

%%%%%%%%%%%%%%%%%%%%%%%%%%%%%%%%%%%%%%%%%%%%%%%%%%%%%%%%%%%%%%%%%%%%%%%%%%%%%%%%%%
We next analyze the general properties of the Josephson current using symmetry arguments.
We first consider the fourfold rotational symmetry $C_{4z}$.
This operation corresponds to a rotation by $\pi/2$ about the $z$ axis and transforms $k_x\rightarrow k_y$，$k_y\rightarrow -k_x$, $\hat{s}_z\rightarrow \hat{s}_z$.
Under this transformation, $M_\mathbf{k}\rightarrow -M_\mathbf{k}$, indicating that $C_{4z}$ reverses the altermagnetic order.
We next consider the time-reversal operator $\mathcal{T}$, which acts as
$k_x\rightarrow-k_x,~k_y\rightarrow-k_y,~\hat{s}_z\rightarrow-\hat{s}_z,~\phi\rightarrow-\phi$.
Combining the two symmetry operations, we define the combined operator $\mathcal{M}=\mathcal{T}C_{4z}$, for which one can show that
\begin{equation}
	\mathcal{M}\hat{\mathcal{H}}_\mathbf{k}(\phi)\mathcal{M}^{-1}=\hat{\mathcal{H}}_\mathbf{k}(-\phi).
\end{equation}
Then, we can conclude that the spectrum of the Hamiltonian $\hat{\mathcal{H}}_\mathbf{k}(\phi)$ satisfies $\{E_n(\phi)\}=\{E_{n^\prime}(-\phi)\}$.
It follows that the free energy $\mathcal{F}$ is an even function of the phase difference $\phi$: $\mathcal{F}(\phi)=\mathcal{F}(-\phi)$.
Therefore, since $I(\phi)\propto \partial\mathcal{F}/\partial\phi$, the Josephson current is an odd function, satisfying $I(\phi)=-I(-\phi)$.
Consequently, the harmonic expansion of the Josephson current contains only sine terms and can be written as
\begin{equation}
	I(\phi)=\sum_{n=1}I_n\sin\left(n\phi\right).
\end{equation}
For later convenience, we express the Josephson current in the normalized form
\begin{equation}
	I(\phi)=I_c\mathcal{I}(\phi),
\end{equation}
where $I_c$ is the critical current and $\mathcal{I}(\phi)$ is the dimensionless current.
The energy of the AMJJ is given by $\mathcal{E}_\mathcal{SAMS}=\int_{0}^t\ V(t^\prime)I[\phi(t^\prime)]dt^\prime$, where $V(t)$ is the voltage across the Josephson junction.
Using the Josephson voltage-phase relation $V(t)=\hbar/(2e)\dot{\phi}$ and evaluating the integral, we obtain the Josephson potential energy of the junction
\begin{equation}
	\mathcal{V}=E_J\mathcal{U}(\phi),
\end{equation}
where the Josephson energy is defined as $E_J=\Phi_0I_c/(2\pi)$, the dimensionless Josephson potential as $\mathcal{U}(\phi)=\int_{-\infty}^\phi \mathcal{I}(\phi^\prime)d\phi^\prime$, and the superconducting flux quantum as $\Phi_0=h/(2e)$.
Here, $h$ and $e$ denote the Planck constant and the elementary charge, respectively.

\begin{figure}
	\centering
	\includegraphics[width=1\textwidth]{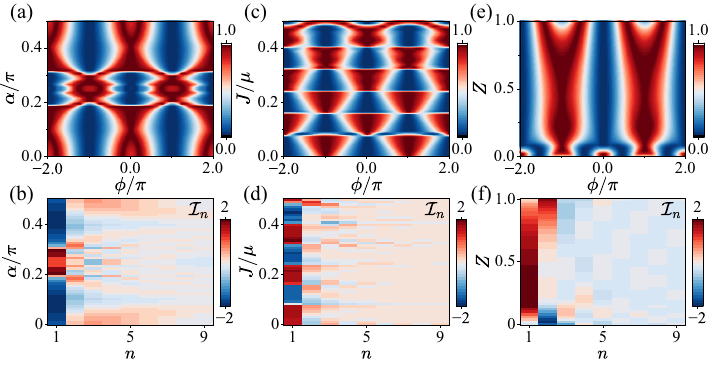}
	\caption{Dependence of the min-max-normalized effective Josephson potential $[\mathcal{V}-\min(\mathcal{V})]/[\max(\mathcal{V})-\min(\mathcal{V})]$ on the junction parameters $\alpha$, $J$, and $Z$. The parameters are listed in Table~\ref{Para_SMFig2}.}
	\label{SMFig2}
\end{figure}

\begin{figure}
	\centering
	\includegraphics[width=1\textwidth]{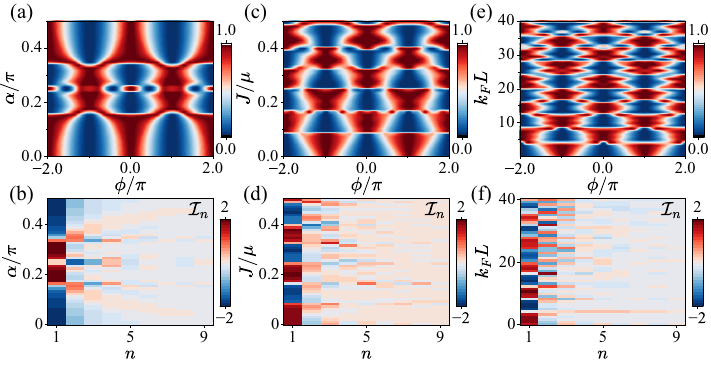}
	\caption{Panels (a)-(d) use the same parameters as Figs.~\ref{SMFig2}(a)-\ref{SMFig2}(d), except that a finite chemical-potential mismatch $\delta\mu/\mu=0.2$ is introduced. Panels (e) and (f) use the same parameters as Figs.~2(a,~b) and 2(c,~d) of the main text, respectively, again with $\delta\mu/\mu=0.2$.}
	\label{SMFig3}
\end{figure}

%%%%%%%%%%%%%%%%%%%%%%%%%%%%%%%%%%%%%%%%%%%%%%%%%%%%%%%%%%%%%%%%%%%%%%%%%%%%%%%%
We next numerically investigate the effects of the microscopic parameters $\alpha$, $L$, $J$, $\delta\mu$, and $Z$ on the Josephson potential.
Figure \ref{SMFig2} shows the effects of $\alpha$, $J$, and $Z$ on the Josephson potential, while those of $L$ and $\delta\mu$ are presented in Fig.~2 of the main text.
Figures \ref{SMFig2}(a)-\ref{SMFig2}(d) show that varying $\alpha$ and $J$ primarily induces $0$-$\pi$ transitions, similar to the effect of $L$ shown in the main text.
In contrast, varying $Z$ enhances the relative contribution of the second harmonic, similar to the effect of $\delta\mu$ shown in the main text.
To illustrate the role of $\delta\mu$ more clearly, we further examine the effects of $\alpha$, $J$, and $L$ on the Josephson potential for a finite value $\delta\mu=0.2\mu$.
A finite $\delta\mu$ enhances the relative contributions of higher-order harmonics to some extent, thereby enlarging the parameter regions supporting the $\phi$- and $2\phi$-junction states (see Figs.~\ref{SMFig3}).
In summary, the Josephson potential can be tailored by tuning these microscopic parameters.

\begin{table*}
	\caption{\label{Para_SMFig2}
		Parameters used in Fig.~\ref{SMFig2}.
	}
	\begin{ruledtabular}
		\begin{tabular}{lccccc}
			\textrm{No.}&
			\textrm{$J/\mu$}&
			\textrm{$\alpha$}&
			\textrm{$\delta\mu/\mu$}&
			\textrm{$k_FL$}&
			\textrm{$Z$}\\
			\colrule
			Fig.~\ref{SMFig2}(a,~b) & $0.4$ & $0\sim\pi/2$ & $0$ & $20$ & $0$ \\
			Fig.~\ref{SMFig2}(a,~b) & $0\sim0.5$ & $\pi/4$ & $0$ & $20$ & $0$ \\
			Fig.~\ref{SMFig2}(e,~f) & $0.4$ & $\pi/4$ & $0.2$ & $20$ & $0\sim 1$\\
		\end{tabular}
	\end{ruledtabular}
\end{table*}

\subsection{Inverse Design}
\subsubsection{Proposed Method}
\begin{figure}
	\centering
	\includegraphics[width=0.35\textwidth]{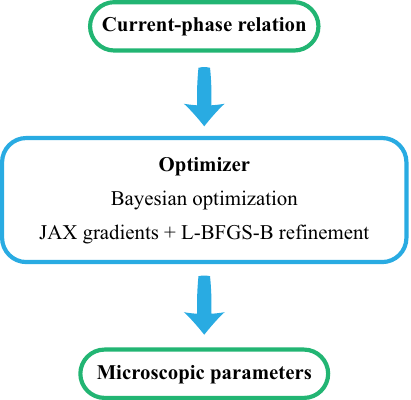}
	\caption{Schematic illustration of the inverse-design workflow.}
	\label{SMFig4}
\end{figure}

\begin{figure}
	\centering
	\includegraphics[width=1\textwidth]{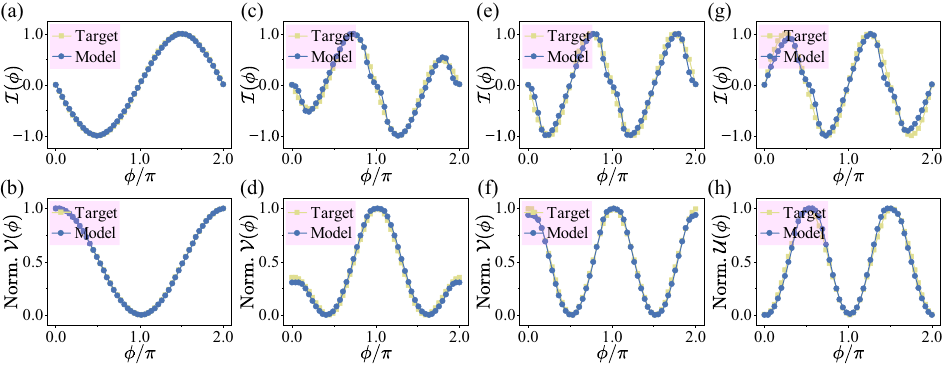}
	\caption{To validate the inverse-design scheme, we consider four distinct target CPRs, whose design parameters are listed in Table~\ref{Para_SMFig5}. Applying the inverse-design workflow shown in Fig.~\ref{SMFig4}, we obtain a set of AMJJ parameters capable of realizing each target CPR, as listed in Table~\ref{Para_SMFig5_opt_Para}.}
	\label{SMFig5}
\end{figure}

As shown above, the numerical results demonstrate that $\pi$, $\phi$, and $2\phi$ junctions can be realized by tuning the microscopic parameters of the AMJJ.
However, owing to the complexity of the microscopic AMJJ model, a closed-form expression relating the CPR to the microscopic parameters is generally unavailable.
We therefore adopt an inverse-design approach to identify the microscopic parameters of the AMJJ that realize the target CPR.
The inverse-design workflow is illustrated in Fig.~\ref{SMFig4}.
Specifically, for a prescribed target CPR, we first employ Bayesian optimization to perform a global search over a broad parameter space.
Once suitable candidate parameters are identified, they are locally refined using a quasi-Newton method, with the required gradients evaluated via automatic differentiation in JAX, yielding a set of microscopic AMJJ parameters that realizes the target CPR.
The loss function is defined as
\begin{equation}
	\mathfrak{L}=\frac{1}{N}\sum_{n=1}^N\left(\mathcal{I}_n-\mathcal{I}_n^T\right)^2,
\end{equation}
where $\mathcal{I}_n^T$denotes the $n$th harmonic coefficient of the target CPR.
To validate the inverse-design method, we consider four representative target CPRs encompassing $\pi$, $\phi$, and $2\phi$ junctions.
The corresponding target parameters are listed in Table~\ref{Para_SMFig5}.
Applying the optimization workflow shown in Fig.~\ref{SMFig4}, we obtain microscopic AMJJ parameters capable of realizing each target CPR.
The corresponding optimization results are shown in Fig.~\ref{SMFig5}, and the optimized parameters are listed in Table~\ref{Para_SMFig5_opt_Para}.
Owing to the highly nonlinear dependence of the AMJJ CPR on the structural parameters, distinct parameters may yield identical or similar CPRs.
Each optimized parameter set reported here represents only one feasible realization of the corresponding target CPR and is therefore not unique.
Figures~\ref{SMFig5}(a) and \ref{SMFig5}(b) correspond to the $\pi$ junction.
Figs.~\ref{SMFig5}(c) and \ref{SMFig5}(d) and Figs.~\ref{SMFig5}(e) and \ref{SMFig5}(f) correspond to two distinct $\phi$ junctions, respectively.
Figures~\ref{SMFig5}(g) and \ref{SMFig5}(h) correspond to the $2\phi$ junction, whose target CPR contains only the second harmonic.
In Fig.~\ref{SMFig5}, the curves with filled-square markers represent the target CPRs, whereas those with filled-circle markers represent the CPRs calculated using the AMJJ structural parameters obtained from the inverse-design procedure.
The two sets of curves are in excellent agreement.
These results validate the proposed inverse-design approach within the microscopic model and indicate that it provides a viable route to engineering the Josephson potential of an AMJJ.

\begin{table*}
	\caption{\label{Para_SMFig5}
		Harmonic components of the target CPRs shown in Fig.~\ref{SMFig5}.
	}
	\begin{ruledtabular}
		\begin{tabular}{lcc}
			\textrm{No.}&
			\textrm{$\mathcal{I}_1^T$}&
			\textrm{$\mathcal{I}_2^T$}\\
			\colrule
			Fig.~\ref{SMFig5}(a,~b) & $-1$ & $0$ \\
			Fig.~\ref{SMFig5}(c,~d) & $0.37$ & $-0.73$ \\
			Fig.~\ref{SMFig5}(e,~f) & $0$ & $-1$ \\
			Fig.~\ref{SMFig5}(g,~h) & $0$ & $1$ \\
		\end{tabular}
	\end{ruledtabular}
\end{table*}

\begin{table*}
	\caption{\label{Para_SMFig5_opt_Para}
		AMJJ parameters obtained via inverse design to realize the target CPRs shown in Fig.~\ref{SMFig5}.
	}
	\begin{ruledtabular}
		\begin{tabular}{lcccccc}
			\textrm{No.}&
			\textrm{$\alpha$}&
			\textrm{$J$}&
			\textrm{$k_FL$}&
			\textrm{$\delta\mu/\mu$}&
			\textrm{$Z$}&
			\textrm{Loss}\\
			\colrule
			Fig.~\ref{SMFig5}(a,~b) & $\pi/4$ & $0.50$ & $4.92$ & $0.43$ & $0.12$ & $0.000035$\\
			Fig.~\ref{SMFig5}(c,~d) & $\pi/4$ & $0.34$ & $19.47$ & $0.22$ & $0.12$ & $0.001164$ \\
			Fig.~\ref{SMFig5}(e,~f) & $\pi/4$ & $0.26$ & $14.62$ & $0.51$ & $0.30$ & $0.001580$\\
			Fig.~\ref{SMFig5}(g,~h) & $\pi/4$ & $0.39$ & $29.04$ & $0.80$ & $0.89$ & $0.003524$\\
		\end{tabular}
	\end{ruledtabular}
\end{table*}

\subsubsection{Effect of Deviations in $\alpha$ on the CPR}
In the preceding discussion, we focus on $\alpha=\pi/4$.
At this orientation, $\hat{\mathcal{M}}_\mathbf{k}=\alpha_1 k_x k_y\hat{\tau}_z\hat{s}_z$.
Because $\hat{\mathcal{M}}_\mathbf{k}$ contains no spin-orbit coupling and the leads are spin-singlet $s$-wave superconductors, the system has a global spin-rotation symmetry.
The N\'eel-vector direction therefore does not affect the current-phase relation.
The parameter $\alpha$ specifies the orientation of the momentum-space form factor relative to the interface normal.
This is a crystallographic quantity fixed by epitaxy, rather than a magnetic one.
This orientation can be controlled: single-variant epitaxy has been demonstrated at wafer scale (RuO$_2$(101) on $r$-plane Al$_2$O$_3$, a single in-plane XRD $\varphi$-scan peak)~\cite{2025HeP82358235}.

\begin{figure}
	\centering
	\includegraphics[width=0.4\textwidth]{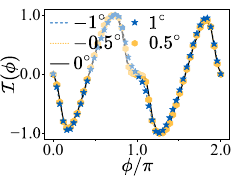}
	\caption{CPRs for angular deviations $\delta\alpha=0^\circ,~\pm0.5^\circ,~\pm1^\circ$ from $\alpha=\pi/4$. The parameters are those listed for Fig.~\ref{SMFig5}(c,~d) in Table~\ref{Para_SMFig5_opt_Para}.}
	\label{SMFigAdd2}
\end{figure}

We now consider the effect of an angular deviation $\delta\alpha$ from $\alpha=\pi/4$ on the CPR.
The actual angle is then $\alpha=\pi/4+\delta\alpha$.
Substituting this angle into the full altermagnet Hamiltonian yields
\begin{equation}
	\hat{H}_\mathrm{AM}\left(\delta\alpha\right)=\frac{\mathcal{J}}{k_F^2}\left[-\left(k_x^2-k_y^2\right)\sin\left(2\delta\alpha\right)+2k_xk_y\cos\left(2\delta\alpha\right)\right]\hat{\tau}_z\otimes\hat{s}_z,
\end{equation}
Under the combined operation $\mathcal{M}_\mathrm{ys}$, defined by $y\rightarrow-y$, $k_y\rightarrow-k_y$, and $s_z\rightarrow -s_z$, we obtain
\begin{equation}
	\mathcal{M}_\mathrm{ys}\hat{H}_\mathrm{AM}\left(\delta\alpha\right)\mathcal{M}_\mathrm{ys}^{-1}=\hat{H}_\mathrm{AM}\left(-\delta\alpha\right).
\end{equation}
The $s$-wave superconductors and the $\delta$ barrier, which is uniform along the interface, are both spin independent and thus invariant under $\mathcal{M}_\mathrm{ys}$.
This yields
\begin{equation}
	\mathcal{M}_\mathrm{ys}\hat{H}_\mathbf{k}\left(\phi,\delta\alpha\right)\mathcal{M}_\mathrm{ys}^{-1}=\hat{H}_\mathbf{k}\left(\phi,-\delta\alpha\right),
\end{equation}
and, in turn,
\begin{equation}
	\mathcal{I}\left(\phi,\delta\alpha\right)=\mathcal{I}\left(\phi,-\delta\alpha\right).
\end{equation}
The numerical results in Fig.~\ref{SMFigAdd2} also confirm this.
The first-order sensitivity of the CPR to $\delta\alpha$ at $\alpha=\pi/4$ is
\begin{equation}
	\left.\frac{\partial \mathcal{I}}{\partial \alpha}\right|_{\alpha=\pi/4}
	=
	\lim_{\delta\alpha\to 0}
	\frac{I(\pi/4+\delta\alpha)-I(\pi/4-\delta\alpha)}{2\delta\alpha}
	=0.
\end{equation}
Thus, the CPR is insensitive to deviations in $\alpha$ to first order.

\section{\label{AMSQ}AM-Based Superconducting Qubit}

\subsection{\label{Tr_H}Hamiltonian}
\begin{figure}
	\centering
	\includegraphics[width=0.4\textwidth]{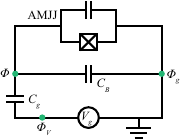}
	\caption{Schematic of a superconducting qubit based on an AMJJ shunted by a capacitor $C_B$.}
	\label{SMFig6}
\end{figure}

As shown in Fig.~\ref{SMFig6}, we consider an AMJJ-based superconducting qubit, in which the AMJJ is shunted by a capacitor $C_B$ and biased by a gate voltage $V_g$ through the gate capacitor $C_g$ .
The AMJJ also contributes a junction capacitance $C_J$.
The green dots indicate the circuit nodes, whose generalized node fluxes are denoted by $\Phi$, $\Phi_g$, and $\Phi_V$, respectively.
For a constant gate voltage $V_g$, the flux of the voltage-source node is $\Phi_V=V_g t$, and that of the ground node is $\Phi_g=0$.
The superconducting phase difference across the AMJJ is $2\pi/\Phi_0(\Phi_g-\Phi)$.
The Hamiltonian of the corresponding AMJJ-based superconducting qubit can then be written as
\begin{equation}
	\hat{H}_\mathrm{Tr}=4E_C\left(\hat{N}-N_g\right)^2+E_J\mathcal{V}(\hat{\phi}).
\end{equation}
Here, $E_C=e^2/(2C_S)$ is the charging energy, $C_S=C_g+C_B+C_J$ is the total capacitance, $N_g=C_gV_g/(2e)$ is the offset charge, and $E_J$ is the Josephson energy.
The flux and phase operators are related by $\hat{\Phi}=\Phi_0/(2\pi)\hat{\phi}$.
In the following, we consider AMJJs whose CPRs are dominated by the first two harmonics, corresponding to one- and two-Cooper-pair tunneling processes, respectively.
The CPR then takes the form
\begin{equation}
	\mathcal{I}\left(\phi\right)=\mathcal{I}_1\sin\left(\phi\right)+\mathcal{I}_2\sin\left(2\phi\right)
\end{equation}
with the corresponding dimensionless Josephson potential given by
\begin{equation}
	\mathcal{V}(\hat{\phi})=-\mathcal{I}_1\cos\left(\hat{\phi}\right)-\frac{\mathcal{I}_2}{2}\cos\left(2\hat{\phi}\right).
\end{equation}

\begin{figure}
	\centering
	\includegraphics[width=0.8\textwidth]{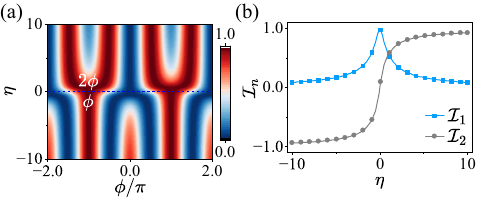}
	\caption{(a) Min–max-normalized effective Josephson potential, $[\mathcal{V}-\min(\mathcal{V})]/[\max(\mathcal{V})-\min(\mathcal{V})]$	for different $\eta$, showing the transition of the AMJJ among $\phi$, $0$, and $2\phi$ junctions. (b) Relative magnitudes of the first and second harmonics, $\mathcal{I}_1$ and $\mathcal{I}_2$, of the Josephson current for different $\eta$.}
	\label{SMFigAdd1}
\end{figure}

As shown in Fig.~\ref{SMFigAdd1}, the AMJJ can transition among the $0$-, $\phi$-, and $2\phi$-junctions as the relative weights of the first and second harmonics are varied.
Specifically, the blue dashed line in Fig.~\ref{SMFigAdd1}(a) denotes the $0$ junction, and the regions above and below it correspond to the $2\phi$ and $\phi$ junctions, respectively.
Figure~\ref{SMFigAdd1}(b) shows how the relative magnitudes of the first and second harmonics, $\mathcal{I}_1$ and $\mathcal{I}_2$, vary with \(\eta\).

\subsection{\label{Tr_CN}Charge Noise}
Charge noise can be broadly classified into high- and low-frequency components.
High-frequency charge noise near the qubit transition frequency induces energy relaxation, whereas low-frequency charge noise primarily causes pure dephasing.

We first consider energy relaxation induced by high-frequency charge noise.
The corresponding relaxation rate can be written as~\cite{2019KrantzP2131821318,2021RasmussenP4020440204}
\begin{equation}
	\Gamma_\mathrm{DC}^Q=\frac{1}{\hbar^2} \left\vert \left\langle \psi_0\left\vert \frac{\partial \hat{H}_\mathrm{Tr}}{\partial N_g}\right\vert \psi_1\right\rangle \right\vert^2 S_Q\left(\omega_q\right).
\end{equation}
Substituting the qubit Hamiltonian into the above expression and simplifying yields
\begin{equation}
	\Gamma_\mathrm{DC}^Q=\frac{1}{\hbar^2} 64E_C^2 \mathcal{N}_\mathrm{DC}^Q S_Q\left(\omega_q\right),
\end{equation}
where $\mathcal{N}_\mathrm{DC}^Q=\vert \langle\psi_0\vert\hat{N}\vert\psi_1\rangle\vert^2$is the transverse charge-noise coupling strength, and $S_Q\left(\omega_q\right)$ denotes the charge-noise spectral density evaluated at the qubit transition frequency $\omega_q$.

%%%%%%%%%%%%%%%%%%%%%%%%%%%%%%%%%%%%%%%%%%%%%%%%%%%%%%%%%%%%%%%%%%%%%%%%%%
Figure~3(c) of the main text shows the transverse charge-noise coupling strength $\mathcal{N}_{\mathrm{DC}}^Q$ as a function of $\eta$ and $E_J/E_C$, where $\eta=\mathcal{I}_2/\mathcal{I}_1$.
For $\eta<0$, the transverse charge-noise coupling strength $\mathcal{N}_{\mathrm{DC}}^Q$ decreases as $\eta$ becomes more negative, indicating a suppression of the qubit sensitivity to charge noise.
In the region $0<\eta<\eta_c$, the transverse charge-noise coupling strength $\mathcal{N}_{\mathrm{DC}}^Q$ increases with $\eta$, indicating an enhanced sensitivity of the qubit to charge noise.
Once $\eta$ exceeds the critical value $\eta_c$, the parity selection rule forces the transverse charge-noise coupling strength $\mathcal{N}_{\mathrm{DC}}^Q$ to vanish, thereby completely suppressing the qubit sensitivity to charge noise.
Here, $\eta_c$ denotes the value of $\eta$ at which the first and second excited states become degenerate, as shown in Fig.~3(a) of the main text.
The small nonzero residual of the transverse charge-noise coupling strength in Fig.~3(c) of the main text results from the numerical error associated with the finite-dimensional discretization of the derivative operator.

\begin{figure}
	\centering
	\includegraphics[width=1\textwidth]{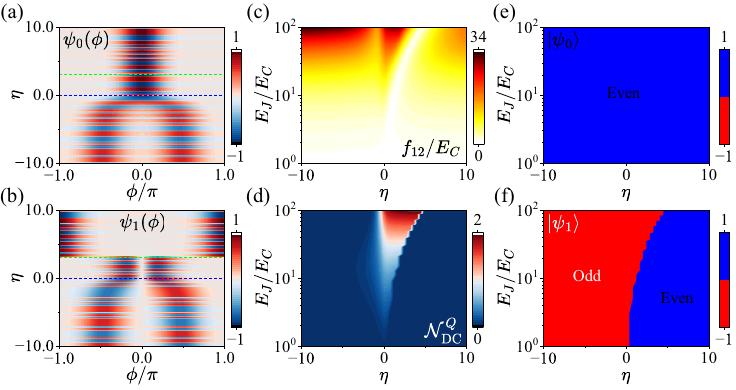}
	\caption{(a,~b) Wave functions of the ground and first excited states in the $\hat{\phi}$ representation for $E_J/E_C=50$. (c) Frequency difference $f_{12}$ between the first and second excited states as a function of $E_J/E_C$ and $\eta$. The curved white band, where $f_{12}\approx0$, indicates the degeneracy of these two states, and its corresponding $\eta$ coordinate defines the critical value $\eta_c$. (d) Transverse charge-noise coupling strength as a function of $E_J/E_C$ and $\eta$. (e,~f) Parities of the ground and first excited states, respectively, over the same parameter range.}
	\label{SMFig7}
\end{figure}

%%%%%%%%%%%%%%%%%%%%%%%%%%%%%%%%%%%%%%%%%%%%%%%%%%%%%%%%%%%%%%%%%%%%%%
The variation in the transverse charge-noise coupling across these parameter regions can be understood from the parities of the qubit eigenstates.
We know that the operator $\hat{N}=-i\partial_\phi$ is an odd-parity operator.
For $\eta<\eta_c$, the ground and first excited states have opposite parities, as shown in Fig.~1(c) of the main text and Fig.~\ref{SMFig7}(a,b).
Thus, the parity selection rule allows a nonzero charge-number-operator matrix element between these states, whose magnitude is determined by their charge-number-weighted overlap in the charge basis.

For $\eta<0$, the charge-number-weighted overlap between the ground and first excited states is small [see Fig.~3(d) of the main text], resulting in a small $\mathcal{N}_\mathrm{DC}^Q$.

For $0<\eta<\eta_c$, the charge-number-weighted overlap between the ground and first excited states increases with $\eta$, leading to an increase in $\mathcal{N}_\mathrm{DC}^Q$.

For $\eta > \eta_c$, the ground and first excited states are both even-parity states, as shown in Fig.~1(e) of the main text and Fig.~\ref{SMFig7}(a,~b).
Because the charge-number operator $\hat{N}$ is an odd-parity operator, its matrix element between the two even-parity states vanishes, $\vert \langle \psi_0\vert\hat{N}\vert \psi_1\rangle\vert^2=0$, thereby completely suppressing energy relaxation induced by transverse charge noise in this region.

As $E_J/E_C$ increases, the critical value $\eta_c$ gradually shifts toward larger $\eta$, as shown in Fig.~3(c) of the main text and Fig.~\ref{SMFig7}(c,~d).
The curved white region with $f_{12}$ in Fig.~\ref{SMFig7}(c) marks the degeneracy between the first and second excited states.
For each $E_J/E_C$, its corresponding horizontal coordinate gives $\eta_c$, which coincides with the boundary of complete suppression of the transverse charge-noise coupling in Fig.~\ref{SMFig7}(d).
Figures~\ref{SMFig7}(e,~f) show the parities of the ground and first excited states, respectively, over the same parameter range.
Throughout the parameter range, the ground state remains even, whereas the first excited state changes from odd to even as $\eta$ crosses $\eta_c$, causing $\mathcal{N}_{\mathrm{DC}}^Q$ to vanish.
This result shows that two-Cooper-pair tunneling enables the qubit to retain substantial anharmonicity while completely suppressing transverse charge-noise coupling, without requiring operation deep in the regime $E_J/E_C\gg1$.

%%%%%%%%%%%%%%%%%%%%%%%%%%%%%%%%%%%%%%%%%%%%%%%%%%%%%%%%%%%%%%%%%%%%%%%
In contrast to high-frequency charge noise, low-frequency charge noise induces pure dephasing of the qubit, with the corresponding dephasing rate given by~\cite{2019KrantzP2131821318,2021RasmussenP4020440204}
\begin{equation}
	\Gamma_\mathrm{DP}^Q=\frac{1}{\hbar^2} \left\vert \left\langle \psi_0\left\vert \frac{\partial \hat{H}_\mathrm{Tr}}{\partial N_g}\right\vert \psi_0\right\rangle - \left\langle \psi_1\left\vert \frac{\partial \hat{H}_\mathrm{Tr}}{\partial N_g}\right\vert \psi_1\right\rangle \right\vert^2 S_Q\left(0\right).
\end{equation}
Substituting the Hamiltonian of the superconducting qubit into the above expression and simplifying yields
\begin{equation}
	\Gamma_\mathrm{DP}^Q=\frac{1}{\hbar^2} 64E_C^2 \mathcal{N}_\mathrm{DP}^Q S_Q\left(0\right),
\end{equation}
where $\mathcal{N}_\mathrm{DP}^Q=\vert \langle\psi_0\vert\hat{N}\vert\psi_0\rangle-\langle \psi_1 \vert\hat{N}\vert \psi_1 \rangle\vert^2$ denotes the longitudinal charge-noise coupling strength and $S_Q(0)$ denotes the charge-noise spectral density at zero frequency.
Likewise, because the Cooper-pair number operator is an odd-parity operator, the parity selection rule enforces $\langle \psi_i\vert\hat{N}\vert\psi_i\rangle=0$.
In other words, pure dephasing of the superconducting qubit induced by low-frequency charge noise is completely suppressed.

\subsection{\label{Tr_FN}Flux Noise without a Superconducting Loop}
Notably, even if a superconducting qubit contains no superconducting loop, time-varying magnetic-field noise may still couple to a single-junction qubit through an induced electromotive force~\cite{KrutiP}.
This effect differs from static frequency tuning caused by loop flux. Its coupling strength depends on the magnetic-field distribution and device geometry~\cite{KrutiP}.
Consider the Hamiltonian of a single-junction altermagnetic qubit
\begin{equation}
	\hat{H}_\mathrm{Tr}=4E_C(\hat{N}-N_g)^2-E_J\mathcal{I}_1\cos(\hat{\phi})-E_J\frac{\mathcal{I}_2}{2}\cos(2\hat{\phi}).
\end{equation}

\begin{figure}
	\centering
	\includegraphics[width=0.8\textwidth]{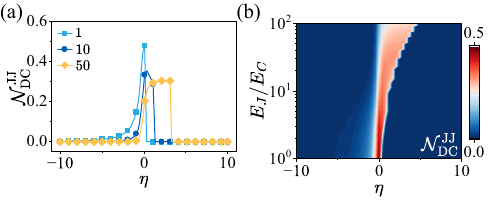}
	\caption{(a) $\mathcal{N}_\mathrm{DC}^\mathrm{JJ}$ as a function of $\eta$ for different values of $E_J/E_C$. (b) Dependence of $\mathcal{N}_\mathrm{DC}^\mathrm{JJ}$ on $E_J/E_C$ and $\eta$.}
	\label{SMFigAdd3}
\end{figure}

Assume that the time-varying magnetic field in the environment appears in the effective circuit model as a common phase perturbation $\chi$ across the junction, and that this perturbation leaves the harmonic coefficients $\mathcal{I}_1$ and $\mathcal{I}_2$ unchanged.
Then, one can get
\begin{equation}
	\hat{H}_\mathrm{Tr}=4E_C(\hat{N}-N_g)^2-E_J\mathcal{I}_1\cos(\hat{\phi}+\chi)-E_J\frac{\mathcal{I}_2}{2}\cos(2\hat{\phi}+2\chi).
\end{equation}
According to Fermi’s golden rule, the relaxation and dephasing rates are respectively
\begin{equation}
	\Gamma_\mathrm{DC}^\mathrm{JJ}=\frac{1}{\hbar^2}E_J^2 \mathcal{N}_\mathrm{DC}^\mathrm{JJ} S_\mathrm{JJ}\left(\omega_q\right)
\end{equation}
and 
\begin{equation}
	\Gamma_\mathrm{DP}^\mathrm{JJ}=\frac{1}{\hbar^2} E_J^2 \mathcal{N}_\mathrm{DP}^\mathrm{JJ} S_\mathrm{JJ}\left(0\right).
\end{equation}
The transverse and longitudinal coupling strengths are respectively
\begin{subequations}
	\begin{align}
		\mathcal{N}_\mathrm{DC}^\mathrm{JJ}&=\left\vert \left\langle \psi_0\left\vert \hat{\mathcal{I}}\right\vert \psi_1\right\rangle \right\vert^2,\\
		\mathcal{N}_\mathrm{DP}^\mathrm{JJ}&=\left\vert \left\langle \psi_0\left\vert \hat{\mathcal{I}}\right\vert \psi_0\right\rangle - \left\langle \psi_1\left\vert \hat{\mathcal{I}}\right\vert \psi_1\right\rangle \right\vert^2,
	\end{align}
\end{subequations}
where $\hat{\mathcal{I}}=\mathcal{I}_1\sin(\hat{\phi})+\mathcal{I}_2\sin(2\hat{\phi})$ is the dimensionless current operator, and $S_\mathrm{JJ}(\omega)$ denotes the noise spectral density.
The above discussion shows that $\hat{\mathcal{I}}$ is an odd-parity operator.
When $\eta$ exceeds $\eta_c$, both the ground state and the first excited state have even parity.
The parity selection rule then gives $\mathcal{N}_\mathrm{DC}^\mathrm{JJ}=0$.
Thus, the relaxation rate due to flux noise coupling to the Josephson current is completely suppressed for $\eta>\eta_c$.
In addition, by the parity selection rule, the dephasing rate due to flux noise coupling to the Josephson current is always zero ($\mathcal{N}_\mathrm{DP}^\mathrm{JJ}=0$).
Fig.~\ref{SMFigAdd3} shows $\mathcal{N}_\mathrm{DC}^\mathrm{JJ}$ as a function of $E_J/E_C$ and $\eta$.
It becomes zero beyond the critical point $\eta_c$.
In the region $\eta<0$ ($\phi$ junction), it is suppressed as $\vert\eta\vert$ increases.

\subsection{\label{Tr_QN}Quasiparticle Noise}
For an $s$-wave superconductor, the isotropic energy gap suppresses energy relaxation induced by quasiparticle-tunneling noise.
The corresponding relaxation rate is given by~\cite{2011CatelaniP6451764517,2011CatelaniP7700277002}
\begin{equation}
	\Gamma_\mathrm{DC}^\mathrm{QP}=\left\vert \left\langle \psi_0\left\vert \sin \left(\frac{\hat{\phi}}{2}\right)\right\vert \psi_1\right\rangle \right\vert^2 S_\mathrm{QP}\left(\omega_q\right),
\end{equation}
where $S_\mathrm{QP}(\omega_q)$ is the quasiparticle-current spectral density at the qubit frequency $\omega_q$.
The spectral density $S_\mathrm{QP}(\omega)$ is defined as~\cite{2011CatelaniP6451764517,2011CatelaniP7700277002}
\begin{equation}
	S_\mathrm{QP}(\omega)=\frac{16E_J}{\pi}e^{-\frac{\Delta}{k_BT}}e^{\frac{2\omega}{2k_BT}}K_0\left(\frac{\vert \omega \vert}{2k_BT}\right),
\end{equation}
where $K_0$ is the modified Bessel function of the second kind.
As can be seen, the isotropic energy gap of the $s$-wave superconductor leads to an exponential suppression of quasiparticle noise, $e^{-\frac{\Delta}{k_BT}}$.
For convenience, we define the squared transition matrix element as $\mathcal{N}_\mathrm{DC}^\mathrm{QP}=\vert \langle \psi_0\vert \sin (\hat{\phi}/2)\vert \psi_1\rangle \vert^2$.

\begin{figure}
	\centering
	\includegraphics[width=0.8\textwidth]{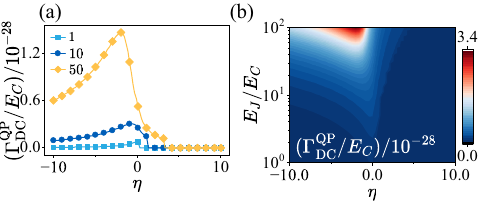}
	\caption{Energy relaxation rate induced by quasiparticle noise: (a) dependence on $\eta$ for different values of $E_J/E_C$; (b) dependence on $E_J/E_C$ and $\eta$.}
	\label{SMFig8}
\end{figure}

%%%%%%%%%%%%%%%%%%%%%%%%%%%%%%%%%%%%%%%%%%%%%%%%%%%%%%%%%%%%%%%%%%%%%%%
In Fig.~\ref{SMFig8}, we present numerical results for the relaxation rate induced by quasiparticle noise.
In Fig.~\ref{SMFig8}(a), we set $E_J/E_C=50$.
As can be seen, after double-Cooper-pair tunneling is introduced, quasiparticle noise is further suppressed with increasing $\vert\eta\vert$.
In particular, when the critical point $\eta_c$ is crossed (the black dashed line), the parity of the first-excited-state wave function changes from odd to even.
Since the ground-state wave function has even parity, whereas the quasiparticle-noise operator $\sin(\hat{\phi}/2)$ is an odd-parity operator, the parity selection rule enforces $\mathcal{N}_\mathrm{DC}^\mathrm{QP}=\vert \langle 0\vert \sin (\hat{\phi}/2)\vert 1\rangle \vert^2=0$.
In other words, quasiparticle noise can be completely suppressed.
In Fig.~\ref{SMFig8}(b), we further examine the effect of varying $E_J/E_C$.
As illustrated in Fig.~\ref{SMFig8}(b), the suppression of quasiparticle noise by double-Cooper-pair tunneling persists regardless of the value of $E_J/E_C$.
However, quasiparticle noise gradually increases with $E_J/E_C$, i.e., as the system moves deeper into the regime $E_J/E_C\gg1$.
In summary, introducing double-Cooper-pair tunneling further suppresses quasiparticle noise in addition to the exponential suppression arising from the energy gap of the $s$-wave superconductor.

\section{\label{FTAMSQ}The Flux-Tunable AM-Based Superconducting Qubit}
To achieve control of the superconducting qubit, we introduce a SQUID composed of two AMJJs.
In this section, we present the derivation of the Hamiltonian and analyze the coherence.
Since the charge-noise analysis is identical to that in Sec.~\ref{Tr_CN}, we focus here on the superconducting qubit’s sensitivity to flux noise and quasiparticle noise.

\subsection{Hamiltonian}
\begin{figure}
	\centering
	\includegraphics[width=0.3\textwidth]{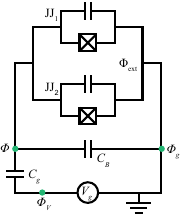}
	\caption{Circuit diagram of the flux-tunable AM superconducting qubit. The qubit is implemented by connecting a SQUID composed of two AMJJs in parallel with a capacitor.}
	\label{SMFig9}
\end{figure}

\begin{figure}
	\centering
	\includegraphics[width=1\textwidth]{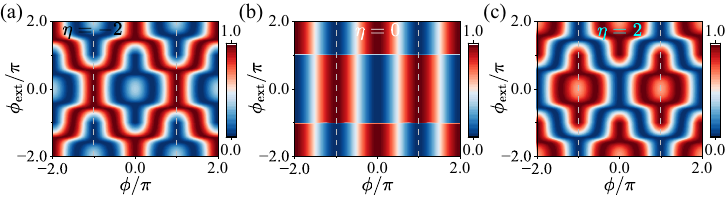}
	\caption{Panels (a-c) show the effective Josephson potential as a function of $\phi_\mathrm{ext}$ for $\eta=-0.4,~0,~0.4$, respectively. Note that we use the min-max-normalized effective Josephson potential $[\mathcal{U}_\mathrm{JJ}^\mathrm{eff}-\min(\mathcal{U}_\mathrm{JJ}^\mathrm{eff})]/[\max(\mathcal{U}_\mathrm{JJ}^\mathrm{eff})-\min(\mathcal{U}_\mathrm{JJ}^\mathrm{eff})]$.}
	\label{SMFig10}
\end{figure}

\begin{figure}
	\centering
	\includegraphics[width=0.8\textwidth]{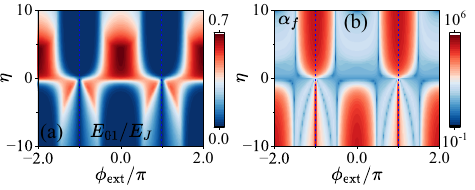}
	\caption{(a) Qubit transition frequency and (b) anharmonicity as functions of the external flux $\phi_\mathrm{ext}$ and the ratio $\eta$. The fixed parameters are $E_J/E_C=50$ and $N_g=0$.}
	\label{SMFigAdd4}
\end{figure}

Figure~\ref{SMFig9}(a) shows a detailed circuit diagram of the flux-tunable AM-based superconducting qubit.
Following the same derivation as in Sec.~\ref{Tr_H}, we obtain the qubit Hamiltonian as
\begin{equation}
	\hat{H}_\mathrm{FTr}=4E_C\left(\hat{N}-N_g\right)^2+\mathcal{U}_\mathrm{JJ}^\mathrm{eff}.
\end{equation}
Here, the effective potential is defined as
\begin{equation}
	\mathcal{U}_\mathrm{JJ}^\mathrm{eff}
	=-E_{1}^{\mathrm{JJ1}}\cos\left(\hat{\phi}-\phi_\mathrm{ext}\right)
	-E_{2}^{\mathrm{JJ1}}\cos\left[2\left(\hat{\phi}-\phi_\mathrm{ext}\right)\right]
	-E_{1}^{\mathrm{JJ2}}\cos\left(\hat{\phi}\right)
	-E_{2}^{\mathrm{JJ2}}\cos\left(2\hat{\phi}\right),
\end{equation}
where
\begin{subequations}
	\begin{align}
		E_{1}^{\mathrm{JJ1}}&=E_{\mathrm{JJ1}}\mathcal{I}_{1}, ~E_{2}^{\mathrm{JJ1}}=E_{\mathrm{JJ1}}\frac{\mathcal{I}_{2}}{2},\\
		E_{1}^{\mathrm{JJ2}}&=E_{\mathrm{JJ2}}\mathcal{I}_{1},~E_{2}^{\mathrm{JJ2}}=E_{\mathrm{JJ2}}\frac{\mathcal{I}_{2}}{2}.
	\end{align}
\end{subequations}
Here, $E_\mathrm{JJi}$ denotes the Josephson energy of the $i$th AMJJ, $E_C=e^2/(2C_S^\mathrm{FTr})$ is the charging energy, and $C_S^\mathrm{FTr}=C_g+C_B+C^\mathrm{JJ1}+C^\mathrm{JJ2}$ is the total capacitance, with $C^\mathrm{JJi}$ denoting the parasitic capacitance of the $i$th AMJJ.
By further rearranging and simplifying the Josephson potential, we obtain
\begin{equation}
	\mathcal{U}_\mathrm{JJ}^\mathrm{eff}=-E_1^S\mathcal{S}_1\left(\phi_\mathrm{ext}\right)\cos\left(\hat{\phi}-\theta_1\right)-E_2^S\mathcal{S}_2\left(\phi_\mathrm{ext}\right)\cos\left(2\hat{\phi}-\theta_2\right),
\end{equation}
where
\begin{subequations}
	\begin{align}
		E_1^S&=E_1^\mathrm{JJ1}+E_1^\mathrm{JJ2},
		~E_2^S=E_2^\mathrm{JJ1}+E_2^\mathrm{JJ2},\\
		\mathcal{S}_1\left(\phi_\mathrm{ext}\right)&=\sqrt{\cos^2\left(\frac{\phi_\mathrm{ext}}{2}\right)+\alpha_1^2\sin^2\left(\frac{\phi_\mathrm{ext}}{2}\right)},\\
		\mathcal{S}_2\left(\phi_\mathrm{ext}\right)&=\sqrt{\cos^2\left(\phi_\mathrm{ext}\right)+\alpha_2^2\sin^2\left(\phi_\mathrm{ext}\right)},\\
		\alpha_1&=\frac{\left\vert E_1^\mathrm{JJ1}-E_1^\mathrm{JJ2}\right\vert}{E_1^S},
		~\alpha_2=\frac{\left\vert E_2^\mathrm{JJ1}-E_2^\mathrm{JJ2}\right\vert}{E_2^S},\\
		\theta_1&=\arctan\left[\alpha_1\tan\left(\frac{\phi_\mathrm{ext}}{2}\right)\right],
		~\theta_2=\arctan\left[\alpha_2\tan\left(\phi_\mathrm{ext}\right)\right].
	\end{align}
\end{subequations}
For convenience, we consider a symmetric junction design, which yields $\alpha_1=\alpha_2=0$.
With this symmetric junction design, the effective Josephson potential simplifies to
\begin{equation}
	\mathcal{U}_\mathrm{JJ}^\mathrm{eff}=E_J\left[-2\mathcal{I}_1\cos\left(\frac{\phi_\mathrm{ext}}{2}\right)\cos\left(\hat{\phi}\right)-\mathcal{I}_2\cos\left(\phi_\mathrm{ext}\right)\cos\left(2\hat{\phi}\right)\right].
\end{equation}
Figures~\ref{SMFig10}(a-c) show the effective Josephson potential as a function of the external flux for $\eta=-0.4,~0,~0.4$, respectively.
The results show that, at $\eta=0$, the Josephson junction supports only single-Cooper-pair tunneling.
Its effective potential can be switched only between the $0$-junction and $\pi$-junction by tuning the external flux $\phi_\mathrm{ext}$.
For $\eta\neq0$, however, the effective potential can realize $0$-, $\pi$-, $\phi$-, and $2\phi$-junctions.
In other words, the relative contributions of single-Cooper-pair and double-Cooper-pair tunneling can be tuned by the external flux.
Figure~\ref{SMFigAdd4} shows that the resonance frequency and anharmonicity of the superconducting qubit can be tuned by external flux.

\subsection{Flux Noise}
Owing to the presence of the SQUID, the superconducting qubit is sensitive to flux noise.
In the main text, we present the dependence of the superconducting qubit’s sensitivity to flux noise on the parameters $\eta$ and $\phi_\mathrm{ext}$.
In this section, we provide the corresponding theoretical derivation.
Low-frequency $1/f$ flux noise induces pure dephasing of the superconducting qubit.
The pure-dephasing rate is given by~\cite{2019KrantzP2131821318,2021RasmussenP4020440204}
\begin{equation}
	\Gamma_\mathrm{DP}^\mathrm{Flux}=\frac{1}{\hbar^2} \left\vert \left\langle \psi_0\left\vert \frac{\partial \hat{H}_\mathrm{FTr}}{\partial \phi_\mathrm{ext}}\right\vert \psi_0\right\rangle - \left\langle \psi_1\left\vert \frac{\partial \hat{H}_\mathrm{FTr}}{\partial \phi_\mathrm{ext}}\right\vert \psi_1\right\rangle \right\vert^2 S_\mathrm{Flux}\left(0\right),
\end{equation}
where $S_\mathrm{Flux}(0)$ denotes the zero-frequency flux-noise spectral density.
Substituting the qubit Hamiltonian into the above expression and simplifying, we obtain
\begin{equation}
	\Gamma_\mathrm{DP}^\mathrm{Flux}=\frac{1}{\hbar^2} E_J^2 \mathcal{N}_\mathrm{DP}^\mathrm{Flux} S_\mathrm{Flux}\left(0\right).
\end{equation}
Here, we define the longitudinal coupling strength to flux noise as
\begin{equation}
	\mathcal{N}_{\mathrm{DP}}^{\mathrm{Flux}}
	=
	\left|
	\begin{aligned}
		&2\mathcal{I}_1\sin\left(\frac{\phi_{\mathrm{ext}}}{2}\right)
		\left[
		\langle\psi_0|\cos(\hat{\phi})|\psi_0\rangle
		-\langle\psi_1|\cos(\hat{\phi})|\psi_1\rangle
		\right]
		\\
		&\quad
		+\mathcal{I}_2\sin\left(\phi_{\mathrm{ext}}\right)
		\left[
		\langle\psi_0|\cos(2\hat{\phi})|\psi_0\rangle
		-\langle\psi_1|\cos(2\hat{\phi})|\psi_1\rangle
		\right]
	\end{aligned}
	\right|^2 ,
\end{equation}
where $\mathcal{F}(x)=-\sin(x)\cos(x)/\vert\cos(x)\vert$.
In Fig.~4(d) of the main text, we present numerical results for $\mathcal{N}_{\mathrm{DP}}^{\mathrm{Flux}}$ as a function of $\eta$ and $\phi_\mathrm{ext}$.
The presence of higher-order harmonics increases the flux-noise sensitivity of the flux-tunable superconducting qubit to some extent.
Nevertheless, this sensitivity can be partially mitigated by operating the qubit near $\phi_\mathrm{ext}=0,~\pm\pi,~\pm2\pi$.

\begin{figure}
	\centering
	\includegraphics[width=0.4\textwidth]{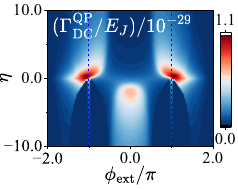}
	\caption{Quasiparticle noise in the flux-tunable AM superconducting qubit as a function of $\eta$ and $\phi_\mathrm{ext}$. Here, we set $E_J/E_C=50$.}
	\label{SMFig11}
\end{figure}

\subsection{Quasiparticle Noise}
The flux threading the SQUID also affects quasiparticle noise.
In the presence of this flux, the energy relaxation rate induced by quasiparticle noise is given by~\cite{2011CatelaniP6451764517,2011CatelaniP7700277002}
\begin{equation}
	\Gamma_{i\rightarrow f}=\sum_{j=1}^M\left\vert \left\langle f\left\vert \sin\left(\frac{\hat{\varphi}_j}{2}\right)\right\vert i\right\rangle \right\vert^2E_{Jj}\widetilde{S}_\mathrm{QP}\left(\omega_{q}\right),
\end{equation}
where $\sin(\hat{\varphi}_j/2)$ is the quasiparticle-tunneling noise operator of the $j$th Josephson junction, $E_{Jj}$ is the Josephson energy of the $j$th Josephson junction, $\omega_{q}$ is the qubit resonance frequency, and $\widetilde{S}_\mathrm{QP}(\omega_{q})=S_\mathrm{QP}(\omega_{q})/E_J$.
Since our model contains only two Josephson junctions, i.e., $M=2$, we obtain
\begin{equation}
	\Gamma_\mathrm{DC}^\mathrm{QP}=E_{J}\mathcal{I}_1\widetilde{S}_\mathrm{QP}\left(\omega_{q}\right)\sum_{j=1,2}\left\vert \left\langle 0\left\vert \sin\left(\frac{\hat{\varphi}_j}{2}\right)\right\vert 1\right\rangle \right\vert^2,
\end{equation}
where $\hat{\varphi}_1=\hat{\phi}-\phi_\mathrm{ext}/2$ and $\hat{\varphi}_2=\hat{\phi}+\phi_\mathrm{ext}/2$.
In Fig.~\ref{SMFig11}, we present numerical results for the energy relaxation rate induced by quasiparticle noise.
Here, we set $E_J/E_C=50$.
As discussed in Sec.~\ref{Tr_QN}, the isotropic energy gap of the $s$-wave superconductor leads to an intrinsic exponential suppression of the superconducting qubit’s sensitivity to quasiparticle noise, reducing $\Gamma_\mathrm{DC}^\mathrm{QP}/E_J$ to a value on the order of $10^{-34}$.
Moreover, double-Cooper-pair tunneling further suppresses the superconducting qubit’s sensitivity to quasiparticle noise.
In particular, once $\eta$ crosses the critical point $\eta_c$ in the $2\phi$-junction regime, quasiparticle noise can be completely suppressed.
In the presence of an external flux $\phi_\mathrm{ext}$, quasiparticle noise can be further suppressed for any value of $\eta$ if $\phi_\mathrm{ext}$ is tuned to a specific range.

% \nocite{*}
% \bibliographystyle{apsrev4-2}
\bibliography{SMReference}